\documentclass[12pt]{article}
\usepackage{geometry}
\usepackage{authblk}
\usepackage{amssymb,stmaryrd}
\usepackage{amsmath}
\usepackage{amsthm}
\usepackage{amsfonts}
\usepackage{amstext}
\usepackage{algorithmic}
\usepackage{xcolor}
\usepackage{algorithm}
\usepackage[labelfont=bf]{caption}
\usepackage{caption}
\usepackage{subcaption}
\usepackage{graphicx}
\usepackage{hyperref}
\theoremstyle{definition}
\newtheorem{exmp}{Example}

\title{Dynamic metabolic resource allocation based on the maximum entropy principle}
\author{David S. Tourigny\thanks{dst2156@cumc.columbia.edu}}
\affil{Columbia University Irving Medical Center\\630 West 168th Street, New York, NY 10032 USA}

\begin{document}
\maketitle

\begin{abstract}
Organisms have evolved a variety of mechanisms to cope with the unpredictability of environmental conditions, and yet mainstream models of metabolic regulation are typically based on strict optimality principles that do not account for uncertainty. This paper introduces a dynamic metabolic modelling framework that is a synthesis of recent ideas on resource allocation and the powerful optimal control formulation of Ramkrishna and colleagues. In particular, their work is extended based on the hypothesis that cellular resources are allocated among elementary flux modes according to the principle of maximum entropy. These concepts both generalise and unify prior approaches to dynamic metabolic modelling by establishing a smooth interpolation between dynamic flux balance analysis and dynamic metabolic models without regulation. The resulting theory is successful in describing `bet-hedging' strategies employed by cell populations dealing with uncertainty in a fluctuating environment, including heterogenous resource investment, accumulation of reserves in growth-limiting conditions, and the observed behaviour of yeast growing in batch and continuous cultures. The maximum entropy principle is also shown to yield an optimal control law consistent with partitioning resources between elementary flux mode families, which has important practical implications for model reduction, selection, and simulation.

\end{abstract}

\section{Introduction}
Dynamic models of metabolism have been introduced as extensions to static, steady state modelling techniques such as flux balance analysis (FBA) \cite{Varma94,Orth10} and elementary flux mode analysis \cite{Schuster94} in order to describe adaptation of cellular activity to changes in the environment. Established examples include dynamic FBA (DFBA) \cite{Mahadevan02}, macroscopic bioreaction models \cite{Provost04,Provost06}, and cybernetic theory based on the optimal control framework by Young and Ramkrishna \cite{Young07,Young08}. Both DFBA and cybernetic theory incorporate regulation of flux across a metabolic reaction network based on some optimality criteria, whereas macroscopic bioreaction models do not and are therefore considered unregulated. Young and Ramkrishna \cite{Young07,Young08} posit that regulatory decisions take the form of a constrained optimisation problem, which must be solved to optimally distribute limited resources among pathways in the network. More recently, various related extensions of DFBA based on resource allocation have been introduced to accommodate the limited capacity for gene expression into dynamic models of metabolism (e.g. \cite{Rugen15,Waldherr15,Lindhorst18}). The concept of resource allocation has also been considered in the static case \cite{Goelzer11,Wortel14,Muller14,Mori16,deGroot19}, where it is suggested that resource constraints arise due to finiteness of the total cellular proteome and the fraction that corresponds to metabolic enzymes. In \cite{Wortel14,Muller14}, it was shown that the FBA solution to the resource allocation problem is to allocate the entirety of resource exclusively to the metabolic pathway maximising the cellular objective.               

From a strategic perspective, cell populations may instead prefer to spread resource among multiple metabolic pathways in order to deal with uncertainty in a fluctuating environment, which could explain the heterogeneity in metabolic pathway use observed experimentally \cite{Levy12,Solopova14,Ackermann15,Martins15,Granados17}. Such `bet-hedging' arguments are akin to various economic theories \cite{Buchen96,Hansen01,Sims03} that posit multiple investments are beneficial to individuals subjected to uncertainty, or that individuals make exclusive investments, but in receipt of slightly different information. These theories are related to the principle of maximum entropy \cite{Jaynes57,Shore80}, because from an information-theoretic standpoint the resource distribution that best represents the current state of knowledge is the one with largest entropy: entropy uniquely satisfies the accepted axioms for an uncertainty measure (up to a constant factor) \cite{Shannon48}, and therefore the maximum entropy distribution consistent with known constraints is uniquely determined as the one that expresses maximum uncertainty with respect to everything else. In biology, this mathematical justification for maximum entropy as an investment strategy that best-accommodates uncertainty forms the basis of various ecological theories (see \cite{Harte14} for a review), and also an interpretation of stem cell multi-potency \cite{Ridden15}. Analogously, in \cite{Kussell05} it was demonstrated that phenotype-switching strategies that are adjusted to the entropy of environmental fluctuations can outperform those that are not. The maximum entropy principle has also recently been applied to static metabolic modelling in various scenarios, including: extensions of FBA to include population heterogeneity \cite{DeMartino17,Fernandez19}, experimental decomposition of fluxes using elementary mode analysis \cite{Zhao09,Zhao10}, and to put forward the suggestion that organisms evolve toward a state of maximum physical entropy \cite{Srienc10,Unrean11}. Hitherto, there has been no attempt to incorporate the maximum entropy principle into dynamic models of metabolism with resource allocation.                 

This paper builds upon the work of Young and Ramkrishna \cite{Young07,Young08} with the purpose of introducing a dynamic model of metabolic resource allocation based on the maximum entropy principle. Although a similar optimal control framework is employed and metabolic network decomposition is also performed using elementary flux modes (EFMs) \cite{Schuster94}, optimality criteria for resource allocation are instead stated in terms of maximum entropy so as to accommodate environmental uncertainty, which produces an original control law. Moreover, the resulting theory is not cybernetic in the sense that there is no reliance on multiple control laws, nor are cybernetic enzymes introduced as auxiliary dynamical variables. The maximum entropy framework unifies DFBA \cite{Mahadevan02} and unregulated macroscopic bioreaction models \cite{Provost04,Provost06} as two limiting extremes of the general theory. A further consequence for dynamic resource allocation is that no assumption beyond maximisation of total catalytic biomass is necessary to describe accumulation of cellular reserve compounds in growth-limiting environments \cite{Reimers17,Tajparast18}. The maximum entropy control also turns out to be consistent with model reduction using EFM families \cite{Song10,Song11,Vilkhovoy16}, which is extremely useful from a modelling point of view because EFM enumeration can result in a combinatorial explosion as metabolic networks grow in size \cite{Klamt02}. 
   
The remainder of this paper is organised as follows: Sections \ref{sec:model} and \ref{sec:control} introduce the dynamic metabolic model and maximum entropy control, which are then extended to include metabolite yields in Section \ref{sec:yields}. Section \ref{sec:families} describes the dynamic maximum entropy framework as applied to model reduction using EFM families, and Section \ref{sec:appplication} presents a specific application of the theory to yeast metabolism. This application is the culmination of a number of working examples found at the end of each section. Additional mathematical details expanding on some parts of the main text can be found in the Appendix.      

\section{Dynamic model of metabolism}
\label{sec:model}
The following dynamical system is considered as a model for metabolism in batch culture
\begin{equation}
\begin{split}
\frac{d}{dt} \mathbf{m}_{ex} &= \mathbf{S}_{ex} \mathbf{v} x  \\
\frac{d}{dt} \mathbf{m}_{in} &= \mathbf{S}_{in}\mathbf{v} - \mu \mathbf{m}_{in} \\
\frac{d}{dt}x &= \mu x , \quad \mu = \mathbf{c}^T \mathbf{v} .
\end{split}
 \label{system}
\end{equation}
Here $\mathbf{m}_{ex}$ ($\mbox{g}\cdot \mbox{L}^{-1}$), $\mathbf{m}_{in}$ ($\mbox{g} \cdot \mbox{L}^{-1} \cdot \mbox{gDW}^{-1} \cdot \mbox{L}$; $\mbox{gDW}$, grams dry weight) are vectors of extra- and intracellular metabolites, respectively, and $\mathbf{S}_{ex}$, $\mathbf{S}_{in}$ the corresponding portions of the stoichiometric reaction matrix $\mathbf{S}$ \cite{Varma94,Orth10,Schuster94}. The scalar variable $x$ ($\mbox{gDW} \cdot \mbox{L}^{-1}$) represents the concentration of total catalytic biomass responsible for catalysing reactions involved in its own production and interconversion of metabolites, and $\mu$ ($\mbox{h}^{-1}$) is the rate of its accumulation (i.e., growth rate) formed as the inner product of the non-negative, $N$-dimensional flux vector $\mathbf{v} = (v_1,v_2,...,v_N)^T$ ($\mbox{g} \cdot \mbox{gDW}^{-1} \cdot \mbox{h}^{-1}$) with the constant coefficient vector $\mathbf{c} = (c_1,c_2,...,c_N)^T$ ($\mbox{gDW} \cdot \mbox{g}^{-1}$). When the reactions are irreversible (which can be assumed after splitting each reversible reaction into two irreversible ones), the reaction fluxes $v_i \geq 0$ can be decomposed as $v_i = e_i f_i(\mathbf{m})$ with $\mathbf{m} = (\mathbf{m}_{ex}, \mathbf{m}_{in})^T$, where $e_i$ is the relative concentration of the enzyme catalysing the $i$th reaction and $f_i(\mathbf{m})$ is the (non-negative) `saturation function' of that enzyme, which includes the thermodynamic driving force, (allosteric) activation or inhibition, and other enzyme-specific effects \cite{deGroot19}. In formulation of this model the relative enzyme concentrations $e_i$ are understood to be control variables, whose values are determined according to control laws in order to satisfy some objective, as are additional arguments of $f_i$ (omitted for notational simplicity) responsible for regulatory effects not directly attributable to the relative level of enzyme $i$. More precisely, while $e_i$ corresponds to relative levels of the $i$th enzyme, its activity is dependent on substrate availability and additional regulatory features, e.g., covalent modification, all encapsulated in the form of a single function $f_i(\mathbf{m})$. It is assumed these two types of control are enacted on distinct time scales so that the $e_i$ are considered {\em slow} control variables while the remaining control variables contained within $f_i(\mathbf{m})$ are considered {\em fast}. For the vast majority of biological models this is a realistic assumption, i.e. the process of transcription and translation of enzyme takes considerably larger than its post-translational regulation (e.g., via phosphorylation).  

Following \cite{Klipp02,deGroot19}, the control variables corresponding to enzyme levels satisfy the constraint 
\begin{equation}
\label{constraint}
\sum_{i=1}^N e_i \leq 1,
\end{equation}
which in this case could correspond to a limited capacity for protein synthesis on ribosomes. In \cite{deGroot19}, expressions like (\ref{constraint}) come with a set of weights or `costs', one for each $e_i$, but here these are absorbed into the $f_i(\mathbf{m})$ (although this is only possible for the case of a single constraint). Regulation of the $e_i$ and remaining fast control variables appearing in (\ref{system}) is assumed to occur such that some metabolic performance index $J$ is maximised, which combined with constraint (\ref{constraint}) introduces the general optimal control problem for resource allocation over the interval $[t_0,t_f]$:
\begin{equation}
\begin{split}
& \mbox{max } J = \Phi_{t = t_f}(\mathbf{m}_{ex}, \mathbf{m}_{in}, x) + \int_{t_0}^{t_f} L(\mathbf{m}_{ex}, \mathbf{m}_{in}, x, \mathbf{e}) dt\\
& \mbox{s.t. (\ref{system}) and (\ref{constraint})} 
\end{split}
\label{problem}
\end{equation}
where $\Phi_{t = t_f}$ is a terminal objective function and $L$ an intermediate objective function evaluated at $\mathbf{e} = (e_1,e_2,...,e_N)^T$. Initial conditions for the dynamic variables in (\ref{system}) may also be given. From a modelling perspective however, it is conventionally not the case that the full system (\ref{system}) is considered due to the immense number of unmeasurable parameters necessary to provide an accurate dynamical description of intracellular metabolism. Common practice is therefore to invoke the {\em quasi-steady state assumption} (QSSA) on intracellular metabolism \cite{Varma94,Schuster94}, which amounts to the assumption that metabolic transients are typically rapid compared to cellular growth rates and changes in the environment. Validity for the QSSA is obtained by comparing the time scale of metabolic processes (fast) to those of transcriptional and translational regulation (slow) \cite{Heinrich96}. Assuming the dilution term $\mu \mathbf{m}_{in}$ is negligible for intracellular metabolites and invoking the QSSA reduces (\ref{system}) to a lower-dimensional system of the form
\begin{equation}
\begin{split}
\frac{d}{dt} \mathbf{m}_{ex} &= \mathbf{S}_{ex} \mathbf{v} x \\
0 &= \mathbf{S}_{in}\mathbf{v}  \\
\frac{d}{dt}x &= \mu  x , \quad \mu = \mathbf{c}^T \mathbf{v} .
\end{split}
 \label{reduced}
\end{equation}

There are two critical issues that should be called into question at this stage. First, the reduction of (\ref{system}) to (\ref{reduced}) based on the QSSA is formal, but it can be rigorously proven that, for fixed relative enzyme concentrations $e_i$, trajectories of the ordinary differential equation (\ref{reduced}) are a good approximation for those of (\ref{system}) provided the conditions of Tikhonov's theorem are met \cite{Khalil02}. These conditions are almost always impossible to validate however, and so typically one needs to assume existence and stability of a quasi-steady state based on biophysical insight. See \cite{Waldherr15} for a discussion of this point. Secondly, and this is not discussed in \cite{Waldherr15}, it is natural to approximate solutions to the optimal control problem (\ref{problem}) using solutions to the reduced problem  
\begin{equation}
\begin{split}
& \mbox{max } J^{red} \\
& \mbox{s.t. (\ref{reduced}) and (\ref{constraint})} ,
\end{split}
\label{redproblem}
\end{equation}
where it is understood that $J^{red}$ is the metabolic performance index evaluated on trajectories of the reduced system (\ref{reduced}). However, to establish validity of this approximation one must appeal to the theory of singularly perturbed optimal control problems \cite{Kokotovic76} and prove that Pontryagin's maximum conditions for the reduced problem (\ref{redproblem}) are equivalent to those obtained by invoking the QSSA on Pontryagin's maximum conditions for the full control problem (\ref{problem}). Unfortunately, establishing equivalence of these two reduction methods remains an open problem for most nonlinear systems. That this equivalence is approximately satisfied should therefore be highlighted as an additional biological assumption for optimal control problems such as those considered here and in \cite{Waldherr15}. The assumption of this equivalence will be referred to as the {\em quasi-reduction equivalent assumption} (QREA).    

Proceeding under the condition that both the QSSA and QREA are valid, a complete set of vectors $\{\mathbf{Z}^k\}_{k=1,2,...,K}$ representing EFMs \cite{Schuster94} or extremal rays for the flux cone $FC= \{ \mathbf{v} : \mathbf{S}_{in} \mathbf{v} = 0 ,  v_i  \geq 0 \quad \forall i \}$ can be used to express any $\mathbf{v} \in FC$ as a conical combination 
\begin{equation}
\label{EFM}
\mathbf{v} = \sum_{k=1}^K \lambda_k \mathbf{Z}^k , \quad \lambda_k \geq 0 \quad k =1,2,...,K.
\end{equation} 
The $\mathbf{Z}^k$ are defined up to some multiplicative constant and the decomposition (\ref{EFM}) represents any $\mathbf{v}$ satisfying constraints imposed by the intracellular component of the stoichiometric matrix \cite{Schuster94}. As reviewed in \cite{Peres18}, there can also be thermodynamic constraints on $\mathbf{v}$, and restricting the set of EFMs to those that satisfy these additional constraints has recently been achieved in \cite{Peres17}. The summation in (\ref{EFM}) may therefore be restricted to a subset of thermodynamically-feasible EFMs, because any thermodynamically-feasible $\mathbf{v}$ can be expressed solely in terms of thermodynamically-feasible EFMs \cite{Jol12}. The converse statement however, that any $\mathbf{v}$ expressed as a linear combination of thermodynamically-feasible EFMs also satisfies the thermodynamic constraints, is not necessarily true, and so this puts a restriction on the interpretation of EFM-based dynamic modelling approaches \cite{{Provost04,Provost06,Song10,Song11,Vilkhovoy16,Baroukh14}}. From the decomposition $v_i = e_i f_i(\mathbf{m})$ one obtains (provided $f_i(\mathbf{m}) \neq 0$)
\begin{equation*} 
e_i = \sum_{k=1}^K \lambda_k \frac{Z^k_i}{f_i(\mathbf{m})} , \quad i =1,2,...,N
\end{equation*} 
where $Z^k_i$ is the $i$th element of the vector representing the $k$th EFM. Therefore constraint (\ref{constraint}) becomes
\begin{equation}
\label{constraint2}
 1 \geq \sum_{k=1}^K \lambda_k \sum_{i=1}^N \frac{Z^k_i}{f_i(\mathbf{m})}  \equiv \sum_{k=1}^K u_k , \quad u_k \geq 0 \quad k =1,2,...,K
\end{equation}
where the new slow control variables $u_k = \lambda_k/r_k$ have been introduced along with 
\begin{equation}
\label{composite}
r_k(\mathbf{m}) =  \left( \sum_{i=1}^N \frac{Z^k_i}{f_i(\mathbf{m})} \right)^{-1} 
\end{equation}
as the `composite' flux through the $k$th EFM (compare with \cite{Young08} in the dynamic and \cite{Wortel14,Muller14} in the static case). 

At this point a choice needs to made for the way that the composite fluxes are to be represented in the reduced system. This is because the QSSA applied to the optimal control problem (\ref{problem}) using the decomposition (\ref{EFM}) to express $\mathbf{v}$ in terms of EFMs does not follow Tikhonov's theorem for ordinary differential equations, which is rather based on determining the slow manifold for $\mathbf{m}_{in}$ in terms of $\mathbf{m}_{ex}$. Common practice is to approximate the functional form of composite fluxes using (e.g., Michaelis-Menten) kinetic rate laws that depend on slow dynamic variables $\mathbf{m}_{ex}$ alone \cite{Provost04,Provost06,Song10,Song11,Vilkhovoy16,Baroukh14}. This choice limits the total number of parameters in the reduced system, but comes with a requirement to select a common normalisation for all EFMs because otherwise the system will not remain invariant to EFM scaling. Such an approximation is made in Section \ref{sec:families} and the application to yeast metabolism presented in Section \ref{sec:appplication}, while for the general discussion it will be assumed that composite fluxes can be well-defined using expression (\ref{composite}) with the fast variables fixed at some constant value, $\mathbf{m}_{in}^*$, independent of the $\mathbf{m}_{ex}$. Substitution for $\mathbf{v}$ in the reduced system (\ref{reduced}) yields 
\begin{equation}
\begin{split}
\frac{d}{dt}\mathbf{m}_{ex} &= x \sum_{k=1}^K r_k(\mathbf{m}_{ex}) \mathbf{S}_{ex} \mathbf{Z}^k u_k \\
\frac{d}{dt}x &= x \sum_{k=1}^K r_k(\mathbf{m}_{ex}) \mathbf{c}^T \mathbf{Z}^k u_k    
\end{split}
 \label{reduced2}
\end{equation}
where only explicit dependence of the $r_k$ on $\mathbf{m}_{ex}$ has been included because the $\mathbf{m}_{in}$ are now assumed constant by the QSSA as stated above. Although the vectors representing EFMs are specified only up to a multiplicative constant, the system (\ref{reduced2}) remains invariant to their re-scaling and is therefore well-defined. Under the QREA one arrives at the reduced optimal control problem
\begin{equation}
\begin{split}
& \mbox{max } J^{red} \\
& \mbox{s.t. (\ref{reduced2}) and (\ref{constraint2})} .
\end{split}
\label{redproblem2}
\end{equation}  
The above form of the dynamic resource allocation problem provides a natural interpretation for each control variable $u_k$ as the fraction of total catalytic biomass concentration $x$ that is allocated to the $k$th EFM. The next section introduces the control law for determining the optimal fraction of this resource.  

\begin{exmp}
\label{example1}
Consider the simplified metabolic network in Figure \ref{fig:1a} as a model for central carbon metabolism, also chosen in \cite{Moller18}. 
\begin{figure}
    \caption{Diagrammatic representations of the simplified metabolic network and corresponding EFMs described in Example \ref{example1}. Arrowheads indicate directionality. Reactions labelled $v_0,v_2,v_3,v_4$ have unit stoichiometry while that labelled $v_1$ has stoichiometry $2$.  $G_{ex}$, $O$, $P_1$, and $P_2$ are extracellular (slow) metabolites whereas $G_{in}$ and $P$ are treated as intracellular (fast) metabolites.}
    \centering
    \begin{subfigure}[t]{\textwidth}
            \caption{Reaction network for central carbon metabolism.} \label{fig:1a}
    \centering
        \includegraphics[width=\linewidth]{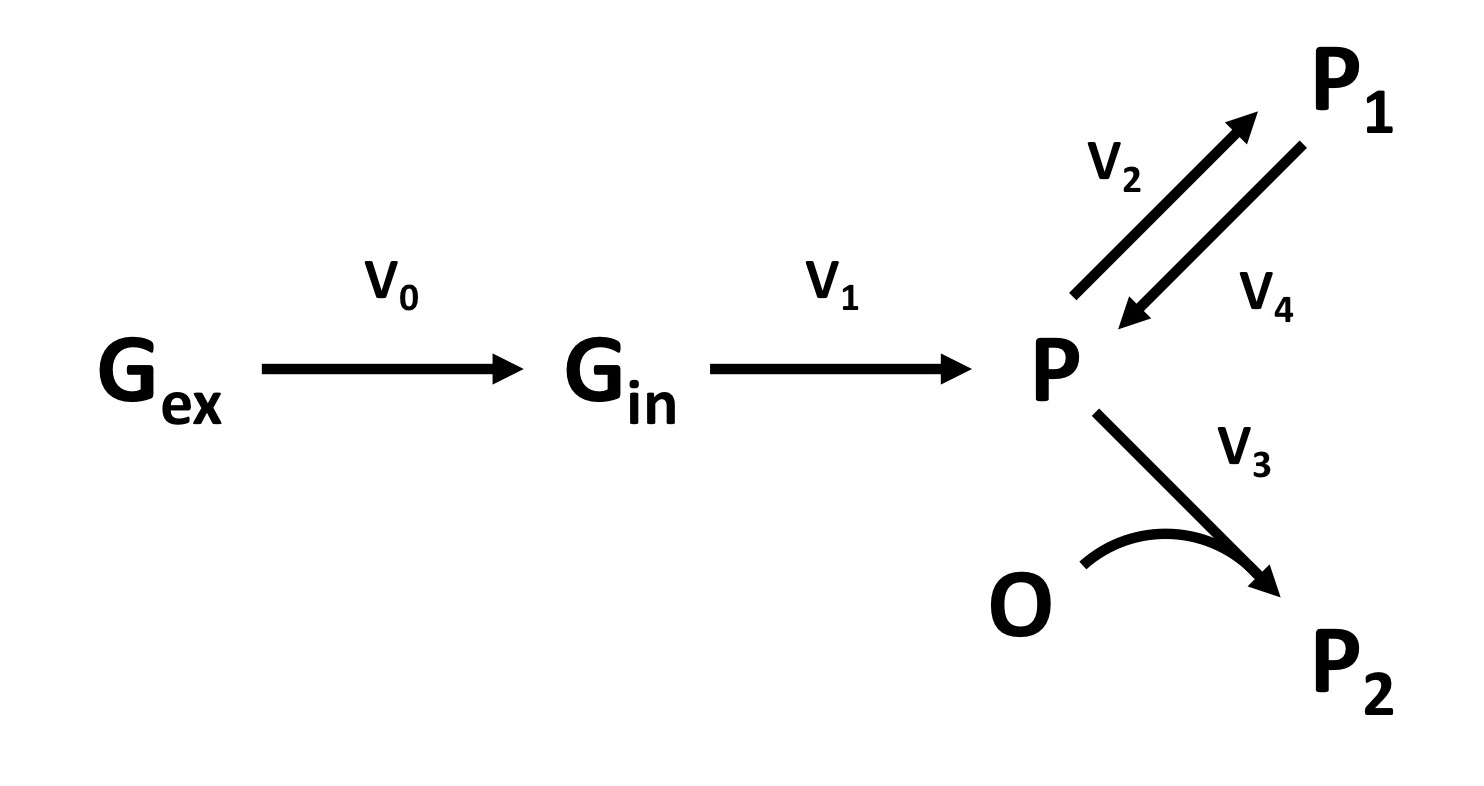} 
    \end{subfigure}
    \vspace{1cm}
    \centering
    \begin{subfigure}[t]{\textwidth}
            \caption{Representation of EFMs by the vectors $\mathbf{Z}^1$, $\mathbf{Z}^2$, $\mathbf{Z}^3$.} \label{fig:1b}
    \centering
        \includegraphics[width=\linewidth]{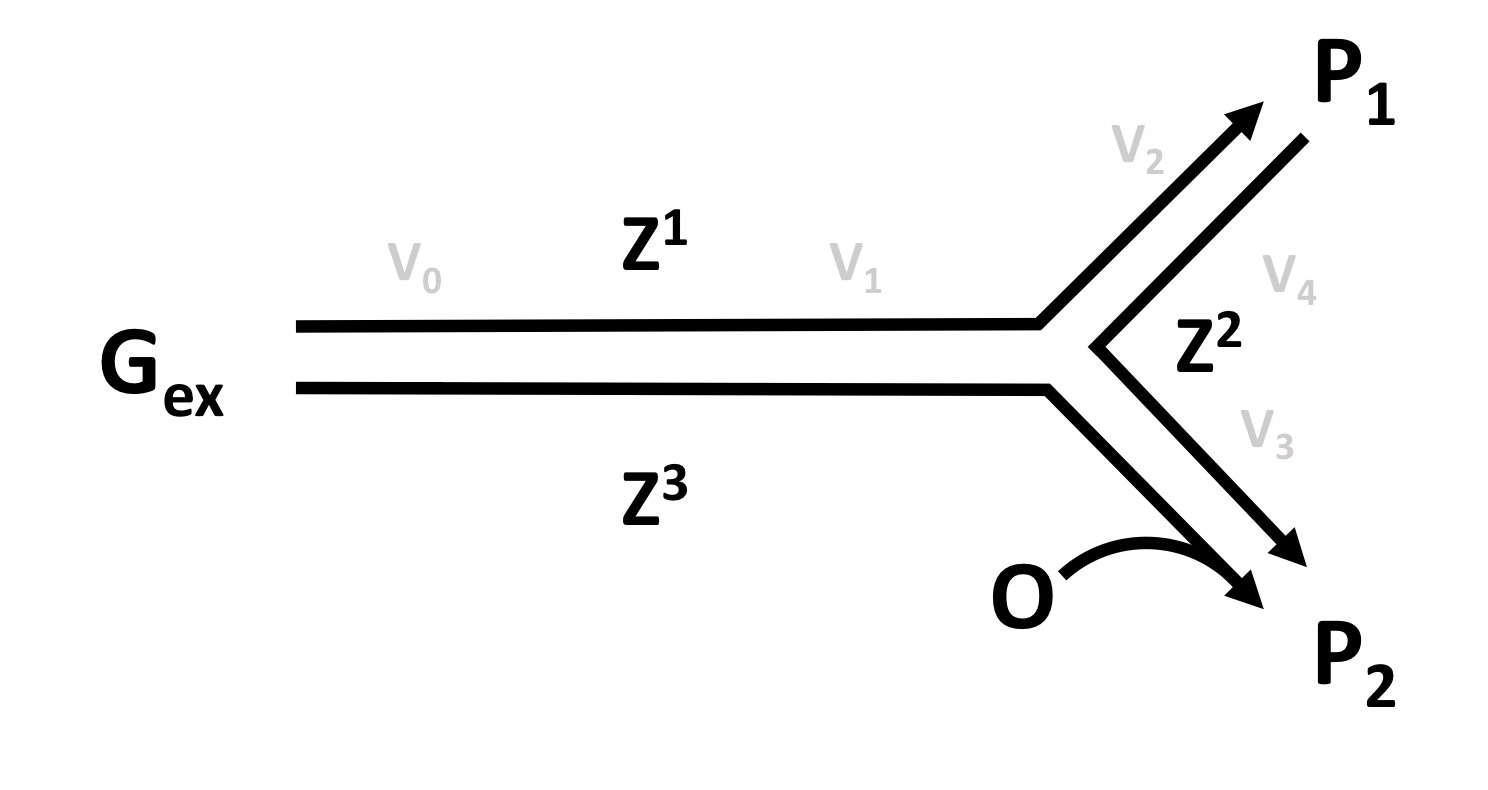} 
    \end{subfigure}
\end{figure}
Concentrations of extracellular metabolites glucose ($G_{ex}$), oxygen ($O$), product 1 ($P_1$), product 2 ($P_2$), and total catalytic biomass ($x$) are slow variables, while concentrations of intracellular glucose ($G_{in}$) and pyruvate ($P$) are fast. In this model only reactions with fluxes $v_1$ and $v_3$ are assumed to contribute directly to the growth rate, such that $c_0 = c_2 = c_4 = 0$ with $c_3>c_1 >0$. Invoking the QSSA on fast intracellular metabolite concentrations, stoichiometric matrices $\mathbf{S}_{ex}$ and $\mathbf{S}_{in}$ give rise to the reduced dynamical system    
\begin{eqnarray*}
\frac{dG_{ex}}{dt} &=& -v_0 x\\
\frac{dO}{dt} &=&   -v_3 x = -\frac{dP_2}{dt} \\
\frac{dP_1}{dt} &=& (v_2 - v_4)x   \\
0 &=& v_0 - v_1   \\
0 &=& 2v_1 - v_2 - v_3 + v_4   \\
\frac{dx}{dt} &=& \mu x , \quad \mu = c_1v_1 + c_3 v_3   .
\end{eqnarray*}
A complete set of three EFMs (represented graphically in Figure \ref{fig:1b}) is provided by the vectors
\begin{equation*}
\mathbf{Z}^1 = \begin{pmatrix}1 \\ 1 \\ 2 \\ 0 \\ 0  \end{pmatrix} , \quad \mathbf{Z}^2 = \begin{pmatrix}0 \\ 0 \\ 0 \\ 1 \\ 1  \end{pmatrix} , \quad \mathbf{Z}^3 = \begin{pmatrix}1 \\ 1 \\ 0 \\ 2 \\ 0 \end{pmatrix} ,
\end{equation*}
and these have respective composite fluxes
\begin{equation*}
r_1(G_{ex}) = \left( \frac{1}{f_0(G_{ex})} + \frac{1}{f_1(G^*_{in})} + \frac{2}{f_2(P^*)} \right)^{-1}
\end{equation*}
\begin{equation*}
r_2(O,P_1) = \left( \frac{1}{f_3(O,P^*)} + \frac{1}{f_4(P_1)}  \right)^{-1}
\end{equation*}
\begin{equation*}
r_3(G_{ex},O) = \left( \frac{1}{f_0(G_{ex})} + \frac{1}{f_1(G^*_{in})} + \frac{2}{f_3(O,P^*)} \right)^{-1}
\end{equation*}
where $G^*_{in}$ and $P^*$ denote the fixed steady state concentrations of intracellular glucose and pyruvate, respectively, and $f_i$ is the saturation function of the $i$th enzyme. Expressed in terms of EFMs and slow control variables $u_1,u_2,u_3$, the reduced dynamical system takes the form
\begin{eqnarray*}
\frac{dG_{ex}}{dt} &=& -[ r_1(G_{ex})u_1 + r_3(G_{ex},O)u_3] x\\
\frac{dO}{dt} &=&   -[r_2(O,P_1)u_2 + 2 r_3(G_{ex},O)u_3] x = -\frac{dP_2}{dt} \\
\frac{dP_1}{dt} &=& [2r_1(G_{ex})u_1 -  r_2(O,P_1)u_2]x   \\
\frac{dx}{dt} &=& [c_1r_1(G_{ex})u_1 + c_3  r_2(O,P_1)u_2 + (c_1 + 2c_3)  r_3(G_{ex},O) u_3]x .
\end{eqnarray*}

\qed
\end{exmp}

\section{Maximum entropy control} 
\label{sec:control}
This section considers the nature of $J^{red}$ in the reduced optimal control problem (\ref{redproblem2}). Related to the separation of timescales for metabolite concentrations in the QSSA arises a similar separation of timescales for control variables. Fast regulatory control variables appearing in $f_i$ are encapsulated within the $r_k(\mathbf{m}_{ex})$, whereas the reduced system (\ref{reduced2}) has a linear dependence on slow control variables $u_k$. In what follows, it will be assumed that fast control variables are selected instantaneously (relative to the QSSA) to yield optimal values of $r_k(\mathbf{m}_{ex})$. On what basis optimality is defined for fast control variables is not of concern, but should derive from biologically reasonable principles. For example, instantaneous maximisation of the composite flux through each EFM amounts to a QSSA-based approximation of the optimal control policy in \cite{Young08} where no separation of timescales was assumed. In this approach, a local objective for each EFM is used to determine optimal values for fast control variables that maximise the composite flux of each EFM individually, and subsequently the slower control variables are chosen to maximise a global objective combining all EFMs. Regardless of the instantaneous policy for selecting fast control variables, the optimal control problem (\ref{redproblem2}) is stated so as to determine the $u_k$ assuming the $f_i$ in $r_k(\mathbf{m}_{ex})$ are given.

Combining a metabolic performance index $J^{red}$ that is linear in the $u_k$ with the constraint (\ref{constraint2}) would result in an optimal control law that allocates the entire fraction of resource exclusively to the EFM with highest return-on-investment \cite{Wortel14,Muller14}. Such a control is the so-called FBA or Bang-Bang policy, which for a variety of evolutionary reasons does not appear to be the most robust nor economically efficient resource allocation strategy in the face of environmental fluctuations \cite{Young07,Solopova14,Ackermann15,Martins15,Granados17,DeMartino17,Fernandez19}. This motivates the revised concept \cite{Young07,Lindhorst18} that regulatory decisions for the control variables $u_k$ should be made based on the projected system response over a (short) time interval of length $\Delta t$. In this sense optimal choices for the $u_k$ are anticipatory of the effects that slower regulatory processes such as transcription and translation will have in the immediate future. Collecting dynamical and control variables into vectors $\mathbf{X}=(\mathbf{m}_{ex},x)^T$ and $\mathbf{u} = (u_1,u_2,...,u_K)^T$, respectively, and writing $\dot{\mathbf{X}} = \mathbf{F}(\mathbf{X},\mathbf{u})$, the linearisation of (\ref{reduced2}) about the state $\mathbf{X}(t)$ and a reference control input $\mathbf{u}^0$ may be assumed a good approximation to the system response at time $t + \tau$ for $\tau \in [0,\Delta t]$ \cite{Young07}. Linearisation yields  
\begin{equation}
\label{linear}
\frac{d}{d\tau} \Delta \mathbf{X} = \mathbf{F}(\mathbf{X}(t),\mathbf{u}^0) + \mathbf{A} \Delta \mathbf{X} + \mathbf{B} \Delta \mathbf{u} 
\end{equation}
where
\begin{equation*}
\mathbf{A} = \frac{\partial}{\partial \mathbf{X}} \mathbf{F}(\mathbf{X}(t),\mathbf{u}^0), \quad \mathbf{B} = \frac{\partial }{\partial \mathbf{u}}\mathbf{F}(\mathbf{X}(t),\mathbf{u}^0) 
\end{equation*}
and $\Delta \mathbf{X}(\tau) = \mathbf{X}(t+\tau) - \mathbf{X}(t)$, $ \Delta \mathbf{u}(\tau) = \mathbf{u}(t+\tau) - \mathbf{u}^0$. When linearising and considering the change in performance index, Young and Ramkrishna \cite{Young07} augmented the accrued benefit derived across during the planning window $[t,t+\Delta t]$ by a term quadratic in the $u_k$ representing the cost or penalty associated with resource allocation. This paper takes a different approach, which is to model the change in performance over the time interval as
\begin{equation}
\Delta J = \mathbf{q}^T \Delta \mathbf{X}(t+\Delta t) + \sigma \int_t^{t+\Delta t}  H(\mathbf{u}) d \tau ,
\end{equation} 
where 
\begin{equation*}
\Delta J = J^{red}(t+\Delta t) - J^{red}(t) , \quad \mathbf{q} = \frac{\partial \phi}{\partial \mathbf{X}} (\mathbf{X}(t)) ,
\end{equation*}
and
\begin{equation}
\label{entropy}
H(\mathbf{u}) = - \sum_{k=1}^K u_k \log (u_k) .
\end{equation}
Here the function $\phi(\mathbf{X})$ represents the metabolic objective of the system and $\sigma$ is a positive parameter that will be interpreted below.

The above choice of $H(\mathbf{u})$ is based on using the maximum entropy principle as a guide for selecting control variables $u_k$, which can be rationalised from several different perspectives: first, since the dynamical model (\ref{reduced2}) is stated in terms of total catalytic biomass of a population, maximum entropy has recently been proposed as an extension of FBA that is intended to capture heterogeneity of different allocation policies adopted by individuals within it \cite{DeMartino17,Fernandez19}. Sources of this heterogeneity include stochasticity in gene expression and phenotype-switching at the single-cell level \cite{Campbell18}, which can serve a functional purpose rather than simply reflecting noise tolerance, and provide a collective advantage to organisms living in fluctuating environments \cite{Levy12,Solopova14,Ackermann15,Martins15,Granados17}. Second, by analogy with decision making problems in finance \cite{Buchen96}, ecology \cite{Harte14}, and communication theory \cite{Johansson05}, distribution of resources according to the principle of maximum entropy is the best choice for maximising expected return-on-investment in the face of uncertainty. As described in the introduction, the maximum entropy principle mathematically captures this notion of bet-hedging because it yields a unique resource allocation strategy consistent with known constraints (e.g., expected return-on-investment given current environmental conditions) while capturing maximum uncertainty in everything else (e.g., future environmental fluctuations) \cite{Jaynes57,Shore80}. Indeed, this second point is intimately tied to the first because population heterogeneity is thought to be one way that cell populations have evolved to execute bet-hedging strategies \cite{Levy12,Solopova14,Ackermann15,Beaumont09}, where both the entropy of the environment \cite{Kussell05} and gene expression profiles \cite{Ridden15} are taken into consideration. Finally, flux decomposition using maximum entropy-weighted EFMs has already been suggested for experimental flux derivation \cite{Zhao09,Zhao10}, or where there is a direct physical interpretation for entropy as that of a chemical reaction \cite{Srienc10,Unrean11}. The former approach uses the maximum entropy principle as it directly applies to model inference \cite{Jaynes57,Shore80}, where uncertainty reflects incompleteness of experimental data and the best statistical model is the one most consistent with those observed. Correspondence of the information-theoretic maximum entropy principle considered here with the physicochemical maximum entropy principle in \cite{Srienc10,Unrean11} are beyond the scope of this paper, but form a deeper relationship between information theory, statistical mechanics, and thermodynamics \cite{Jaynes57}.       

It is reasonable to assume that biological systems evolve under selection for maximal fitness by exploiting the capability to fully utilise their resource (the same assumption is made in \cite{Young07}). This implies the total summation constraint in (\ref{constraint2}) is satisfied as an exact equality and, because of the remaining non-negativity constraints, the vector $\mathbf{u}$ of resource fractions now can be interpreted as a discrete probability distribution across the EFMs. Applying Pontryagin's maximum principle to the optimal control problem
\begin{equation}
\label{linprob}
\begin{split}
& \mbox{max } \Delta J \\
& \mbox{s.t. (\ref{linear}) and } \sum_{k=1}^K u_k = 1, \quad u_k \geq 0 \quad k=1,2,...,K
\end{split}
\end{equation}                
and setting $\tau = 0$ as explained in Appendix \ref{maxent}, results in the following alternative to the optimal control provided in \cite{Young07}:
\begin{equation}
\label{control}
u_k(t) = \frac{1}{Q}\exp \left( \frac{1}{\sigma} \mathbf{q}^T \mathbf{e}^{\mathbf{A} \Delta t} \mathbf{B}^k\right) .
\end{equation}  
Here $\mathbf{B}^k$ denotes the $k$th column of $\mathbf{B}$, $\mathbf{e}^{\mathbf{A} \Delta t}$ is the matrix exponential of $\mathbf{A} \Delta t$, and the normalisation factor $Q$ is the partition function
\begin{equation*}
Q = \sum_{k=1}^K \exp \left( \frac{1}{\sigma} \mathbf{q}^T \mathbf{e}^{\mathbf{A} \Delta t} \mathbf{B}^k \right) . 
\end{equation*}
As described by Jaynes in \cite{Jaynes57}, the control (\ref{control}) is the Boltzmann distribution with $\sigma$ taking the place of temperature and {\em effective return-on-investment}
\begin{equation}
\mathcal{R}^k_{\Delta t} = \mathbf{q}^T \mathbf{e}^{\mathbf{A} \Delta t} \mathbf{B}^k
\label{eroi}
\end{equation}
for the $k$th EFM taking the place of energy. Use of the adjective `effective' will become clear shortly. In the limit $\sigma \to 0$, the control (\ref{control}) collapses to the Bang-Bang/FBA policy \cite{Wortel14,Muller14} where all resource is allocated to the EFM with the greatest effective return-on-investment (\ref{eroi}) (although this does imply the $u_k$ can change rapidly, whereas formally they should be treated as slow control variables). Conversely, $u_k \to 1/K$ $(\forall k = 1,2,..., K)$ as $\sigma$ grows so that resource is partitioned equally among all EFMs in the limit $\sigma \to \infty$. This indifferent distribution of resource among EFMs is equivalent to the unregulated macroscopic bioreaction models of Provost and Bastin \cite{Provost04,Provost06}. Clearly neither extreme is necessarily an ideal representation of the optimal regulatory process, and therefore $\sigma > 0$ is taken to be finite so that the resource is allocated amongst EFMs according to their effective return-on-investment (larger getting more). What proportional majority of resource is awarded to the EFM with greatest effective return-on-investment is determined by the precise value of $\sigma$, which is considered to be a parameter fine-tuned over the course of evolution.

The general definition of effective return-on-investment (\ref{eroi}) depends on a specific choice of metabolic objective that throughout the remainder of this paper is assumed to be maximisation of total catalytic biomass, i.e. $\phi(\mathbf{X}) = x$, which results in $\mathbf{q}=(\mathbf{0},1)^T$. The vector $\mathbf{B}^k$ is obtained by evaluating the derivative of $\mathbf{F}$ with respect to $u_k$ at $\mathbf{X}(t)$, and since $\mathbf{F}$ is linear in $u_k$ this choice of $\phi$ results in     
\begin{equation}
\label{bk}
\mathbf{B}^k = x r_k(\mathbf{m}_{ex}) \begin{pmatrix}  \mathbf{S}_{ex}  \\ \mathbf{c}^T  \end{pmatrix}  \mathbf{Z}^k .
\end{equation}
In the first instance it is assumed that only immediate consequences of the injected control actions need to be considered when evaluating $\mathbf{u}$, and therefore $\Delta t = 0$ (the next section will consider non-zero choices of $\Delta t$ that involve additional complexity due the matrix exponential of $\mathbf{A} \Delta t$). By analogy with \cite{Young07}, when $\Delta t = 0$, the control law (\ref{control}) will be termed the {\em greedy maximum entropy control}. This simplifying assumption, that the Jacobian matrix $\mathbf{A}$ does not appear in the effective return-on-investment (\ref{eroi}), is mathematically equivalent to the biological statement that future changes in the environment are not taken into consideration when making regulatory decisions. As described in Section \ref{sec:yields}, higher order corrections to the effective return-on-investment could be accounted for by a biological mechanism that has evolved to anticipate such environmental changes, but with $\Delta t = 0$ the greedy maximum entropy control serves to maximise expected return-on-investment given the current state of the environment but complete uncertainty about the future. Using the greedy maximum entropy control, the effective return-on-investment for the $k$th EFM reduces to
\begin{equation}
\label{return}
\mathcal{R}^k_{0}(\mathbf{m}_{ex}) = xr_k(\mathbf{m}_{ex}) \mathbf{c}^T \mathbf{Z}^k \equiv x \mbox{R}_0^k(\mathbf{m}_{ex})
\end{equation}   
where notation $\mbox{R}_0^k(\mathbf{m}_{ex})$ has been introduced for the return-on-investment evaluated at zeroth order ($\Delta t = 0$). Multiplication of $\mbox{R}_0^k(\mathbf{m}_{ex})$ by $x$ gives the {\em greedy effective return-on-investment} $\mathcal{R}^k_{0}(\mathbf{m}_{ex})$. Just as in the case of system (\ref{reduced2}), zeroth-order return-on-investment $\mbox{R}_0^k(\mathbf{m}_{ex})$ and the corresponding optimal control remain invariant to re-scaling of $\mathbf{Z}^k$ because this is cancelled by the same factor appearing in the composite flux $r_k$ (\ref{composite}). The greedy effective return-on-investment (\ref{return}) for the $k$th EFM is therefore proportional to a weighted harmonic mean of the $f_i(\mathbf{m}_{ex})$ multiplied by a weighted arithmetic mean of the $c_i$. The weighting for the $k$th EFM is provided by the $N$ components $Z^k_i$ and the conclusion is that the greatest proportion of resource is allocated to the EFM for which the product of these two means is the largest. 

In contrast to the resource allocation rules obtained by Young and Ramkrishna \cite{Young07}, observe that the greedy maximum entropy control law (\ref{control}) implies {\em all} EFMs, including those with with zero or negative zeroth-order return-on-investment, will be allocated a non-zero fraction of resource provided $x/\sigma$ remains finite. Spreading of resource between multiple pathways is known to be optimal for dealing with uncertainty in a non-deterministic environment \cite{Solopova14,Ackermann15,Martins15,Granados17,Kussell05,DeMartino17,Fernandez19}, and investing in each EFM is a bet-hedging strategy analogous to those in behavioural economics \cite{Hansen01,Sims03} that captures remaining uncertainty when it is not possible to anticipate future environmental conditions. Equipped only with knowledge about the current environment, allocating a small fraction of resource (e.g. fraction of the proteome) to EFMs not contributing directly to growth is not considered wasteful because there is always a small probability that one of these pathways will have a benefit in the future \cite{OBrien16}. Higher order corrections to return-on-investment will be described in Section \ref{sec:yields}, but without additional information the greedy maximum entropy control law spreads the remaining fraction of resource indiscriminately between EFMs with zero zeroth-order return-on-investment. The remaining resource fraction will tend to be very small when a majority of resource is heavily concentrated on EFMs having large return-on-investment (that are relatively more likely to be of benefit), combined with the appearance of total catalytic biomass $x$ as an overall scaling factor in (\ref{return}). As $x$ increases, it plays an opposing role to $\sigma$ in the control law (\ref{control}), meaning the distribution of resources amongst EFMs will become more heavily concentrated on those yielding the greatest return-on-investment (i.e., optimal resource allocation approaches the Bang-Bang/FBA policy as $x \to \infty$ with $\sigma$ fixed). This observation aligns well with the suggestion that the maximum entropy distribution represents the cumulative behaviour of individuals within a (finite) population \cite{DeMartino17,Fernandez19}, since the spread of the population distribution will tend to decrease as the number of individuals within it increases. 

\begin{exmp}
\label{example2}
The greedy effective return-on-investments for the three EFMs in Example \ref{example1} are given by
\begin{eqnarray*}
\mathcal{R}_0^1(G_{ex}) &=& xc_1 r_1(G_{ex}) \\
\mathcal{R}_0^2(O,P_1) &=& xc_3 r_2(O,P_1)    \\
\mathcal{R}_0^3(G_{ex},O) &=& x(c_1 + 2c_3) r_3(G_{ex},O) .
\end{eqnarray*}
For illustrative purposes, assume that $r_k(\mathbf{m}_{ex}) \to 1$ ($k=1,2,3$) as all extracellular metabolite concentrations become saturating. This implies that when $G_{ex}$, $O$, and $P_1$ are very large the greedy maximum entropy control law gives
\begin{equation*}
u_1 \approx \frac{1}{Q}\exp(xc_1/\sigma) , \quad u_2 \approx \frac{1}{Q}\exp(xc_3/\sigma), \quad u_3 \approx \frac{1}{Q}\exp(x(c_1+2c_3)/\sigma)
\end{equation*}
where $Q = e^{xc_1/\sigma} + e^{xc_3/\sigma} + e^{x(c_1+2c_3)/\sigma}$. Since $c_3 > c_1$, in this case the EFM represented by $\mathbf{Z}^3$ receives the greatest fraction of resource, followed by that represented by $\mathbf{Z}^2$, and finally the EFM represented by $\mathbf{Z}^1$ receives the smallest fraction. On the other hand, if $O$ becomes very small while $G_{ex}$ and $P_1$ remain saturating, i.e., oxygen concentrations become limiting, instead $\mathcal{R}_0^1 >>  \mathcal{R}_0^2 \approx \mathcal{R}_0^3$ and in this case the majority of resource is allocated to the EFM represented by $\mathbf{Z}^1$. Conversely, if glucose concentrations $G_{ex}$ become limiting while $O$ and $P_1$ remain saturating, this results in  $\mathcal{R}_0^2 >>  \mathcal{R}_0^1 \approx \mathcal{R}_0^3$ and the majority of resource is allocated to the EFM represented by $\mathbf{Z}^2$.

\qed
\end{exmp}

\section{Metabolite yields and anticipatory regulation}
\label{sec:yields}
In previous sections, $\mathbf{m}_{ex}$ was used to denote the concentrations of extracellular metabolites, assuming that all intracellular metabolites are considered {\em fast} and therefore approximately constant at any instantaneous moment in time by the QSSA. This neglected the possibility that certain intracellular metabolites may not satisfy the QSSA criteria and instead vary on the {\em slow} timescale associated with $\mathbf{m}_{ex}$ and $x$. Examples of slowly varying intracellular metabolites include storage compounds (see recent work \cite{Rugen15,Reimers17,Tajparast18}), which are suggested to increase growth rate across a time interval that includes several diverse environmental extremes, e.g., a 24h day-night epoch or feast-famine cycle. Such rationale may explain the regulation of storage pathways in organisms found in environments with predictable dynamics, but fails to describe the general patterns of accumulation and utilisation outside of this regime. As a relevant example, consider the case of intracellular carbohydrate reserves \cite{Holme57,Preiss89,Lillie80,Francios01}. The observed accumulation of intracellular carbohydrates in response to nutrient limitation is not intuitively rationalised based on choosing a metabolic objective of maximising total catalytic biomass alone, because investing resources in any process not contributing directly to growth would be considered a sub-optimal control policy. For this reason, among others, authors have considered alternative metabolic objectives, such as maximising total carbon uptake, or have explicitly included intracellular reserves as an integral component of biomass \cite{Feist10,Waldherr17,Lakshmanan19}. However, here it is demonstrated that no further assumption beyond $\phi(\mathbf{X}) = x$ (the metabolic objective of maximising total catalytic biomass described in Section \ref{sec:control}) is necessary for explaining the accumulation of storage compounds in response to nutrient limitation. The discussion also involves evaluating the optimal control law (\ref{control}) with $\Delta t > 0$, which by analogy with \cite{Young07} is called the {\em temporal maximum entropy control}.                                  

To make the exposition more concrete, it will be useful to distinguish between two types of EFMs as suggested in Section \ref{sec:control}: those that contribute directly to growth, such that $\mathbf{c}^T \mathbf{Z}^k > 0$; and those that do not, such that $\mathbf{c}^T \mathbf{Z}^k = 0$. The case $\mathbf{c}^T \mathbf{Z}^k < 0$ is excluded from consideration, but this is not a particularly restrictive assumption because in the vast majority of models all $c_i$ will be non-negative as are, necessarily, all vector components $Z^k_i$. Typical control laws based on the choice $\phi(\mathbf{X}) = x$, like the FBA/Bang-Bang policy and the greedy control law of Young and Ramkrishna \cite{Young07}, preclude the allocation of resources to EFMs with $\mathbf{c}^T \mathbf{Z}^k = 0$ since then $\mathcal{R}_0^k(\mathbf{m}_{ex}) = 0$ also. These control policies therefore neglect possible benefits of allocating resources to EFMs contributing to processes other than growth directly, such as accumulation of storage compounds that may be utilised for growth should environmental conditions become unfavourable. In Section \ref{sec:control}, it was shown that the greedy maximum entropy control allocates a fraction of resource to every EFM, including those with $\mathbf{c}^T \mathbf{Z}^k = 0$, which accounts for maximal uncertainty when only information about the current environment is available. Correspondingly, the fraction of resource allocated to EFMs with $\mathbf{c}^T \mathbf{Z}^k = 0$ will tend to increase as the average effective return-on-investment of EFMs with $\mathbf{c}^T \mathbf{Z}^k > 0$ decreases. On the other hand, the Jacobian matrix $\mathbf{A}$ appears in the effective return-on-investment (\ref{eroi}) of the temporal maximum entropy control, which is equivalent to the biological statement that regulatory decisions also take into consideration effects that the control action will have on the environment in the immediate future. If, for example, a system has evolved to anticipate that formation of storage compounds provides a future opportunity to increase total catalytic biomass, this further reduction in uncertainty is accommodated into the temporal maximum entropy control. One consequence is that individual EFMs not contributing directly to growth can receive greater (or less) investment should they involve consumption or production of metabolites that make the future environment more (or less) favourable for growth. This is most clearly demonstrated by understanding the higher order corrections to return-on-investment that arise for small $\Delta t > 0$.

When $\Delta t$ is small, the matrix exponential $\mathbf{e}^{A\Delta t}$ can be approximated to first order so that the effective return-on-investment (\ref{eroi}) becomes $\mathcal{R}^k_{\Delta t}(\mathbf{m}_{s}) \approx x[R_0^k(\mathbf{m}_s) + \Delta t R_1^k(\mathbf{m}_s)]$, where the first-order correction to return-on-investment derived in Appendix \ref{correction} is 
\begin{equation}
\label{froi}
R^k_1(\mathbf{m}_{s}) =  \bar{R}_0(\mathbf{m}_{s}) R^k_0(\mathbf{m}_{s}) + x r_k(\mathbf{m}_{s}) \left( \frac{\partial \bar{R}_0}{\partial \mathbf{m}_{s}} (\mathbf{m}_{s}) \right)^T \mathbf{S}_s \mathbf{Z}^k .
\end{equation}  
Here $\mathbf{m}_{s}$ is used to denote all slow metabolite concentrations with corresponding stoichiometric matrix $\mathbf{S}_s$ and
\begin{equation*}
\bar{R}_0(\mathbf{m}_{s}) = \sum_{k=1}^K  R^k_0(\mathbf{m}_{s}) u^0_k
\end{equation*}
is an average of the zeroth-order return-on-investment (i.e., the average contribution to growth rate) provided by components of the reference control $\mathbf{u}^0$ at time $t$. Observe that when the reference control is taken to be the uniform one (as suggested by Young and Ramkrishna \cite{Young07}), corresponding to the $\sigma \to \infty$ limit of the maximum entropy control, then $\bar{R}_0(\mathbf{m}_{s})$ is simply the arithmetic mean of the $R^k_0(\mathbf{m}_{s})$. There are two terms in the first-order correction to return-on-investment (\ref{froi}): the first is the product $\bar{R}_0(\mathbf{m}_{s}) R^k_0(\mathbf{m}_{s})$, which is always non-negative and obviously large when both $\bar{R}_0(\mathbf{m}_{s})$ and $R^k_0(\mathbf{m}_{s})$ are large; the second term is 
\begin{equation}
\label{troi1}
x r_k(\mathbf{m}_{s})  \left( \frac{\partial \bar{R}_0}{\partial \mathbf{m}_{s}} (\mathbf{m}_{s}) \right)^T \mathbf{S}_s \mathbf{Z}^k \equiv x Y_k(\mathbf{m}_{s}) .
\end{equation}
To understand the newly defined quantity $Y_k(\mathbf{m}_{s})$, note that for any slow metabolite concentration $m$ one has 
\begin{equation}
\label{deriv}
\frac{\partial R^k_0}{\partial m}(\mathbf{m}_s) = \mathbf{c}^T\mathbf{Z}^k  \frac{ \partial r_k}{\partial m}(\mathbf{m}_s)  .
\end{equation}
Vector components of the form (\ref{deriv}), one for each slow metabolite, provide a measure of how dependent the average contribution to growth rate is on concentration $m$ at time $t$. If $r_k$ is monotonically increasing (which is true if all $f_i$ are monotonically increasing) then each component is non-negative and a large value of (\ref{deriv}) indicates that a change in $m$ leads to a relatively large increase of $\bar{R}_0(\mathbf{m}_{s})$, i.e., $m$ is growth-limiting at time $t$; conversely, a value close to zero indicates a change in the concentration of that metabolite has a negligible effect, i.e., $m$ is not growth-limiting at time $t$. In general, it will not always be true that the $f_i$ are monotonically increasing, in which case some of the components (\ref{deriv}) can be negative indicating certain slow metabolite concentrations may be growth-prohibiting. In either case, values (\ref{deriv}) serve to weight components of the vector $\mathbf{S}_s \mathbf{Z}^k$, which can be either positive or negative because they provide the yield of each metabolite for the $k$th EFM. A positive yield indicates the $k$th EFM will contribute to the production of a metabolite, while a negative yield means the EFM will contribute to its consumption. $Y_k(\mathbf{m}_{s})$ as defined in (\ref{troi1}) is therefore interpreted as the {\em total metabolite yield} for the $k$th EFM, with weighting of each individual metabolite yield proportional to the relative ability of the corresponding metabolite to increase $\bar{R}_0(\mathbf{m}_{s})$. The relative sizes of $\bar{R}_0(\mathbf{m}_{s}) R^k_0(\mathbf{m}_{s})$ and $xY_k(\mathbf{m}_{s})$ determine whether the first-order correction $R^k_1(\mathbf{m}_{s})$ is positive or negative. A positive first-order correction to the return-on-investment on implies $\mathcal{R}^k_{\Delta t}(\mathbf{m}_{s}) > \mathcal{R}^k_0(\mathbf{m}_{s})$ whereas a negative correction means that $\mathcal{R}^k_{\Delta t}(\mathbf{m}_{s}) < \mathcal{R}^k_0(\mathbf{m}_{s})$. 

Consequences of using the temporal maximum entropy control can then be summarised as follows: when no slow metabolites are growth-limiting or growth-prohibiting, the average contribution to growth rate $\bar{R}_0(\mathbf{m}_s)$ is large relative to magnitudes of the $xY_k(\mathbf{m}_s)$, and therefore resource becomes further concentrated on EFMs with $\mathbf{c}^T \mathbf{Z}^k>0$. However, when one or more slow metabolite is growth-limiting, and consequently the average contribution to growth rate $\bar{R}_0(\mathbf{m}_s)$ is low, EFMs with non-negative total metabolite yield $Y_k(\mathbf{m}_s)$ (such as those with $\mathbf{c}^T\mathbf{Z}^k = 0$) can be allocated a larger fraction of resource than in cases where $\bar{R}_0(\mathbf{m}_s)$ is high. This type of behaviour has been observed in most microbial populations \cite{Holme57,Preiss89,Lillie80,Francios01}. For example, the storage carbohydrate glycogen is produced by yeast upon limitations in extracellular carbon or nitrogen, and in bacteria glycogen accumulates under conditions of limiting growth when carbon is in excess but other nutrients are deficient (see \cite{Wilson10} for a review). Also in yeast, up-regulation of trehaolse synthesis is known to serve as an indicator for cell populations with lower growth rates \cite{Levy12}, which has been rationalised using a bet-hedging argument. The greedy maximum entropy control generates such an inverse correlation between average contribution to growth rate and the levels of activation of EFMs with $\mathbf{c}^T\mathbf{Z}^k = 0$, but only the temporal maximum entropy control distinguishes between them based on total metabolite yields and their ability to shape environmental conditions. In conclusion, both maximum entropy control laws account for accumulation of intracellular reserves under growth-limiting conditions without imposing any assumption on the objective other than maximisation of total catalytic biomass. However, the temporal maximum entropy control law describes some form of anticipatory regulation, whereas the greedy maximum entropy control law accommodates maximal uncertainty if only current environmental conditions are known.   

\begin{exmp}
\label{example3}
Consider the simplified metabolic network in Figure \ref{fig:2a} as an extension of the one introduced in Example \ref{example1}. 
\begin{figure}
    \caption{Extension of the metabolic network described in Example \ref{example1} to include a storage compound $C$ treated as an additional slow metabolite. The new reactions labelled $v_5,v_6$ both have unit stoichiometry and give rise to three additional EFMs as described in Example \ref{example3}.}
    \centering
    \begin{subfigure}[t]{\textwidth}
            \caption{Reactions of central carbon metabolism with storage.} \label{fig:2a}
    \centering
        \includegraphics[width=\linewidth]{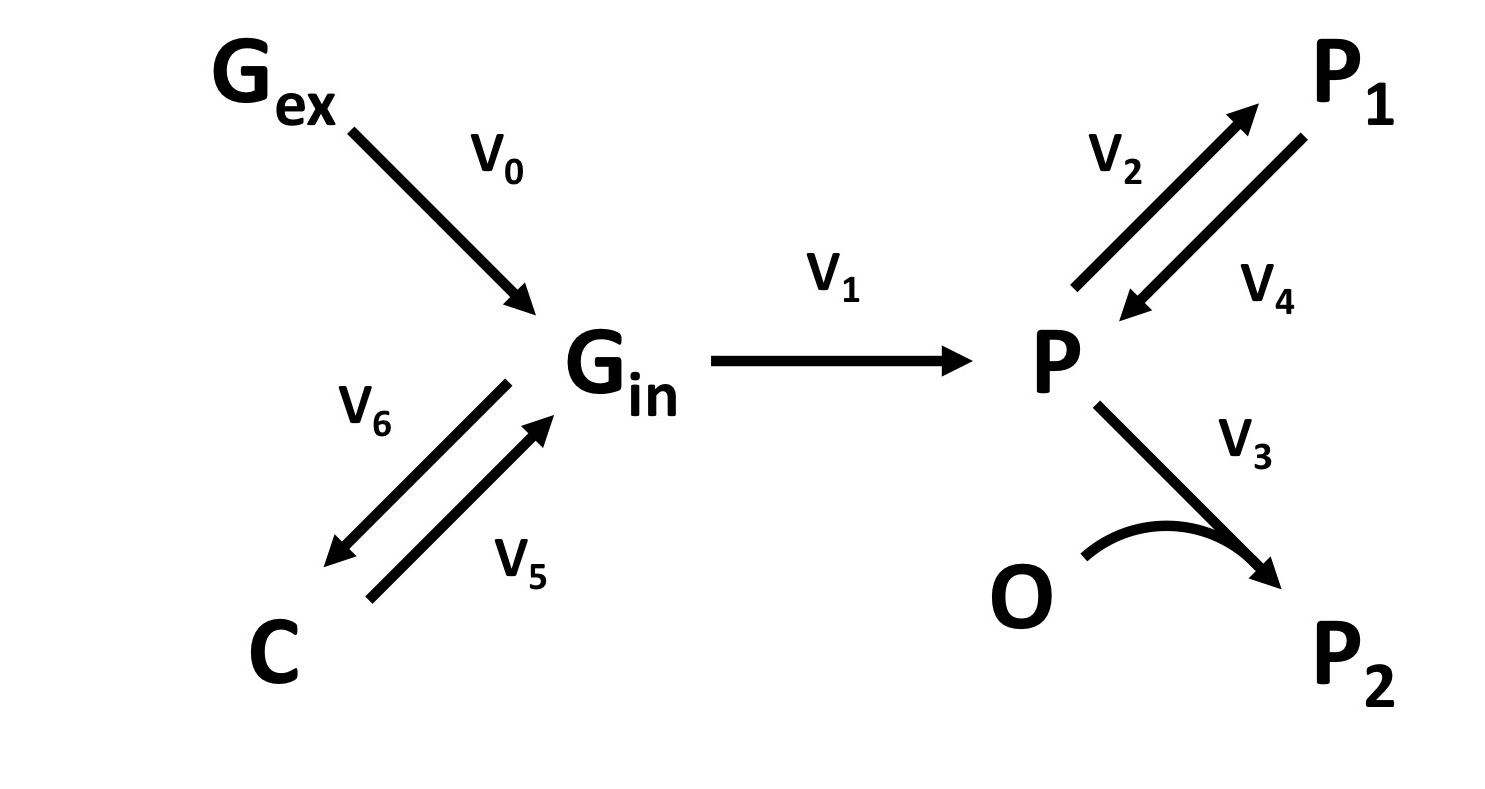} 
    \end{subfigure}
    \vspace{1cm}
     \begin{subfigure}[t]{0.45\textwidth}
     \centering
     \includegraphics[width=\linewidth]{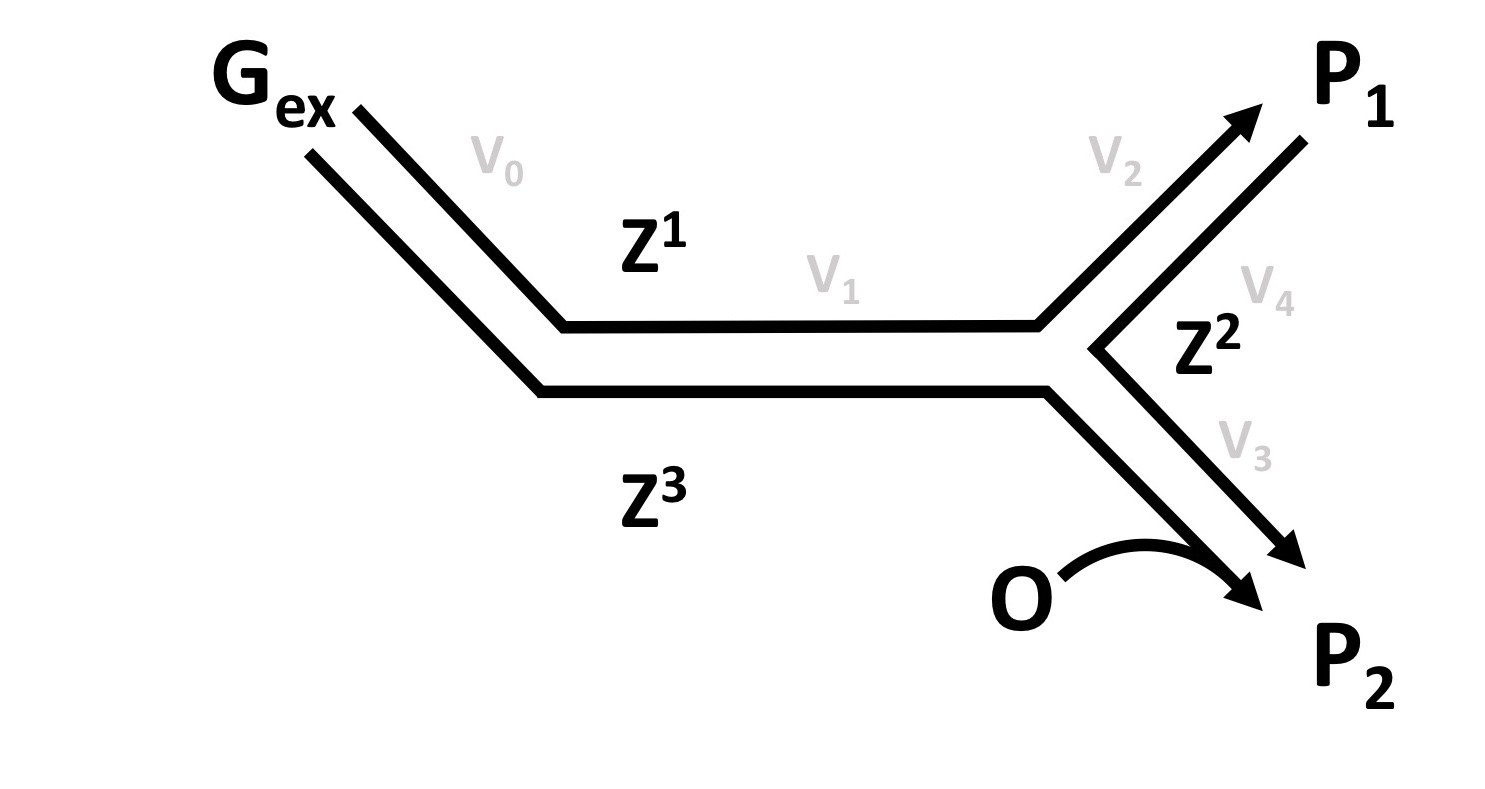} 
     \caption{Representation of EFMs by the vectors $\mathbf{Z}^1$, $\mathbf{Z}^2$, $\mathbf{Z}^3$.} \label{fig:2b}
    \end{subfigure}
    \hfill
    \begin{subfigure}[t]{0.45\textwidth}
        \centering
        \includegraphics[width=\linewidth]{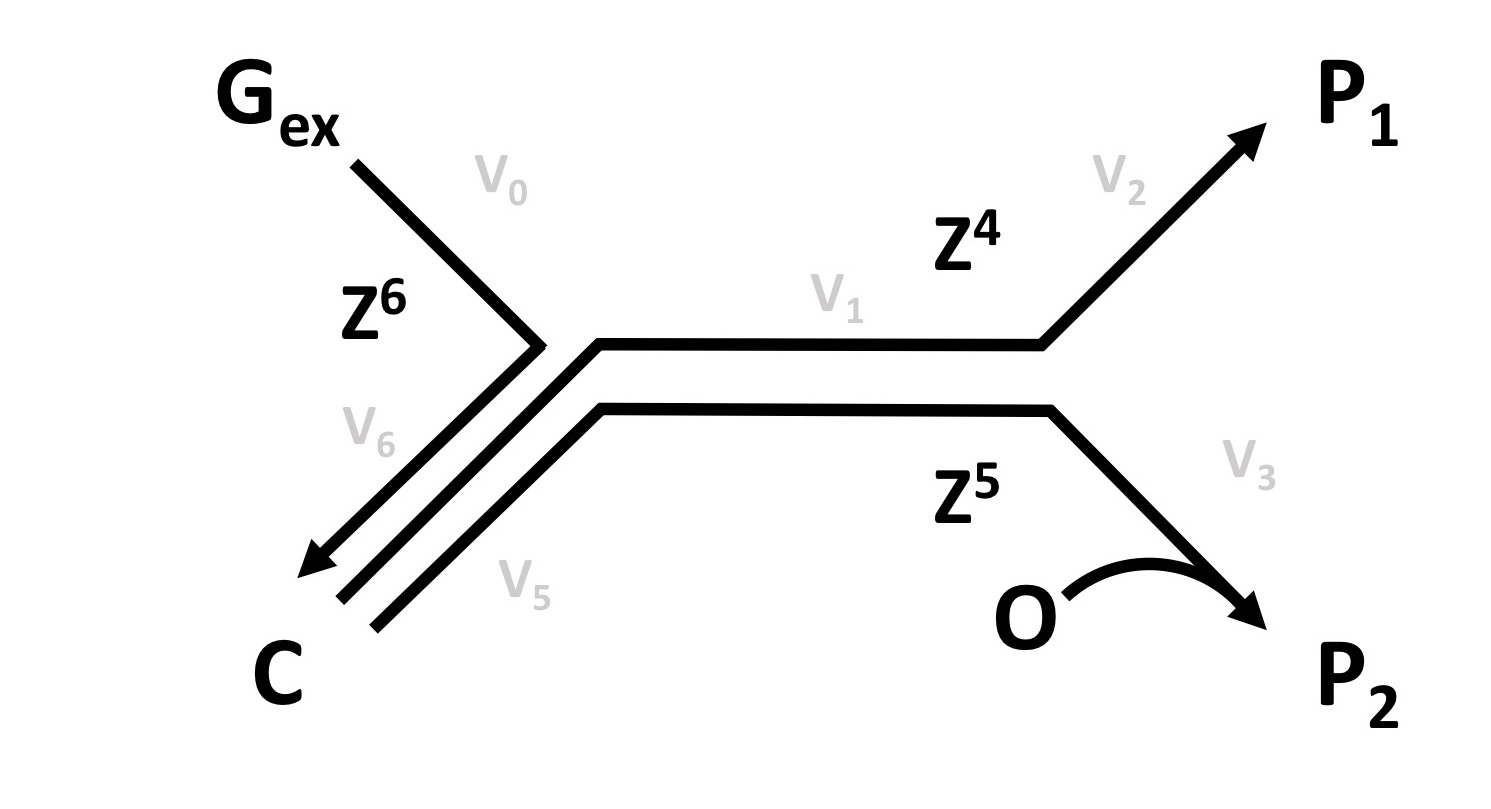} 
        \caption{Representation of EFMs by the vectors $\mathbf{Z}^4$, $\mathbf{Z}^5$, $\mathbf{Z}^6$.} \label{fig:2c}
    \end{subfigure}
\end{figure}
The storage compound with concentration $C$ is introduced as a slow intracellular metabolite. Its consumption and production imply the addition of two reactions to the network with fluxes $v_5$ and $v_6$, respectively, which do not to contribute directly to growth (i.e., $c_5 = c_6 = 0$). The reduced system from Example \ref{example1} is extended to include an additional dynamical term for the new slow variable
\begin{equation*}
\frac{dC}{dt} = (v_6 - v_5) - \mu C
\end{equation*}   
and the algebraic equations arising from the QSSA are modified to  
\begin{eqnarray*}
0 &=&  v_0 -v_1 + v_5 - v_6\\
0 &=& 2v_1 -v_2 -v_3 +v_4  .
\end{eqnarray*}
A complete set of six EFMs (represented graphically in Figures \ref{fig:2b} and \ref{fig:2c}) is provided by the vectors
\begin{equation*}
\mathbf{Z}^1 = \begin{pmatrix}1 \\ 1 \\ 2 \\ 0 \\ 0 \\ 0 \\ 0 \end{pmatrix} , \quad \mathbf{Z}^2 = \begin{pmatrix}0 \\ 0 \\ 0 \\ 1 \\ 1 \\ 0 \\ 0  \end{pmatrix} , \quad \mathbf{Z}^3 = \begin{pmatrix}1 \\ 1 \\ 0 \\ 2 \\ 0 \\ 0 \\ 0 \end{pmatrix} , \quad \mathbf{Z}^4 = \begin{pmatrix}0 \\ 1 \\ 2 \\ 0 \\ 0 \\ 1 \\ 0  \end{pmatrix} , \quad \mathbf{Z}^5 = \begin{pmatrix}0 \\ 1 \\ 0 \\ 2 \\ 0 \\ 1 \\ 0 \end{pmatrix} , \quad \mathbf{Z}^6 = \begin{pmatrix}1 \\ 0 \\ 0 \\ 0 \\ 0 \\ 0 \\ 1 \end{pmatrix} .
\end{equation*}
The composite EFM fluxes $r_1,r_2,r_3$ and greedy effective return-on-investments $\mathcal{R}_0^1,\mathcal{R}_0^2,\mathcal{R}_0^3$ are those given in Examples \ref{example1} and \ref{example2}, respectively, while
 \begin{equation*}
r_4(C) = \left( \frac{1}{f_1(G^*_{in})} + \frac{2}{f_2(P^*)} + \frac{1}{f_5(C)} \right)^{-1}
\end{equation*}
\begin{equation*}
r_5(O,C) = \left(\frac{1}{f_1(G^*_{in})} + \frac{2}{f_3(O, P^*)} + \frac{1}{f_5(C)}  \right)^{-1}
\end{equation*}
\begin{equation*}
r_6(G_{ex}) = \left( \frac{1}{f_0(G_{ex})} + \frac{1}{f_6(G^*_{in})} \right)^{-1} 
\end{equation*}
and
\begin{equation*}
\mathcal{R}_0^4(C) = x c_1 r_4(C), \quad \mathcal{R}_0^5(O,C) = x(c_1 + 2c_3) r_5(O,C) , \quad \mathcal{R}_0^6 = 0 .
\end{equation*}
Omitting explicit dependencies of the $r_k$ on slow metabolites to ease notation, the reduced system expressed in terms of EFMs and control variables $u_k$ is 
\begin{eqnarray*}
\frac{dG_{ex}}{dt} &=& -[ r_1u_1 + r_3u_3 + r_6u_6] x\\
\frac{dO}{dt} &=&   -[r_2u_2 + 2 r_3u_3 + 2 r_5u_5] x = - \frac{dP_2}{dt} \\
\frac{dP_1}{dt} &=& [2r_1u_1 -  r_2u_2 + 2r_4u_4]x   \\
\frac{dC}{dt} &=& r_6u_6 -  r_4u_4 - r_5u_5 - \mu C   \\
\frac{dx}{dt} &=& [c_1(r_1u_1 + r_4u_4)  + c_3  r_2u_2 + (c_1 + 2c_3) ( r_3u_3 + r_5u_5)]x.
\end{eqnarray*}
The metabolite yields for each EFM are supplied by the vectors
\begin{equation*}
\mathbf{S}_s \mathbf{Z}^1 = \begin{pmatrix}-1 \\ 0 \\ 2 \\ 0  \end{pmatrix} , \quad \mathbf{S}_s \mathbf{Z}^2 = \begin{pmatrix}0 \\ -1 \\ -1 \\ 0   \end{pmatrix} , \quad \mathbf{S}_s \mathbf{Z}^3 = \begin{pmatrix} -1 \\ -2 \\ 0 \\ 0  \end{pmatrix} ,
\end{equation*}
\begin{equation*}
\mathbf{S}_s \mathbf{Z}^4 = \begin{pmatrix}0 \\ 0 \\ 2 \\ -1   \end{pmatrix} , \quad \mathbf{S}_s \mathbf{Z}^5 = \begin{pmatrix}0 \\ -2 \\ 0 \\ -1  \end{pmatrix} , \quad \mathbf{S}_s \mathbf{Z}^6 = \begin{pmatrix}-1 \\ 0 \\ 0 \\ 1 \end{pmatrix} ,
\end{equation*}
and, assuming for simplicity that the oxygen concentration $O$ is saturating at time $t$ so that $\partial r_k(t) /\partial O =0 $ ($k = 1,2,...,6$), the total metabolite yields evaluated using the uniform reference control $u^0_k = 1/6$ ($k = 1,2,...,6$) are
\begin{eqnarray*}
Y_1(t) &=& - \frac{r_1}{6}\left[c_1 \frac{\partial (r_1 + r_3)}{\partial G_{ex}} + 2c_3 \left( \frac{\partial r_3}{\partial G_{ex}} -  \frac{\partial r_2}{\partial P_1} \right) \right]_t \\
Y_2(t) &=& - \frac{r_2}{6}\left[c_3 \frac{\partial r_2}{\partial P_1} \right]_t \\
Y_3(t) &=&  - \frac{r_3}{6}\left[ c_1 \frac{\partial (r_1 + r_3)}{\partial G_{ex}} + 2c_3 \frac{\partial r_3}{\partial G_{ex}} \right]_t \\
Y_4(t) &=& - \frac{r_4}{6}\left[  c_1 \frac{\partial (r_4 + r_5) }{\partial C}  + 2c_3 \left( \frac{\partial r_5}{\partial C} -  \frac{\partial r_2}{\partial P_1} \right)   \right]_t \\
Y_5(t) &=& - \frac{r_5}{6}\left[ c_1 \frac{\partial (r_4+r_5)}{\partial C} + 2c_3 \frac{\partial r_5}{\partial C} \right]_t \\
Y_6(t) &=& \frac{r_6}{6}\left[ c_1 \left( \frac{\partial r_4}{\partial C} - \frac{\partial r_1 }{\partial G_{ex}} \right) + (c_1 + 2c_3) \left( \frac{\partial r_5 }{\partial C} - \frac{\partial r_3}{\partial G_{ex}} \right)\right]_t
\end{eqnarray*}
where $[\cdot]_t$ indicates the expression inside square parentheses is to be evaluated at $t$. Observe that in a regime where metabolite concentrations are such that
\begin{equation*}
\frac{\partial r_2(t)}{\partial P_1} < \frac{\partial r_3(t)}{\partial G_{ex}} < \frac{\partial r_5(t)}{\partial C} \quad \mbox{and} \quad \frac{\partial r_1(t)}{\partial G_{ex}} < \frac{\partial r_4(t)}{\partial C} ,
\end{equation*} 
then $Y_k(t) < 0$ for $k = 1,2,...,5$ whereas $Y_6(t)>0$. In fact, when oxygen is not saturating it can be shown that the contributions from non-zero derivatives $\partial r_k(t) /\partial O$ ($k=2,3,5$) decrease $Y_2,Y_3,Y_5$ further while leaving $Y_1,Y_4,Y_6$ unchanged.

\qed

\end{exmp}  

\section{Model reduction using EFM families} 
\label{sec:families}
This section explores the practical aspects of model design and simulation. Enumeration of EFMs for large stoichiometry matrices $\mathbf{S}$ can lead to a combinatorial explosion as their number grows with increasing network size and connectivity \cite{Klamt02}. This necessitates inclusion of many undetermined parameters and control variables in the reduced system (\ref{reduced2}), which are difficult to model accurately should sufficient experimental data not be available. Consequently, previous attempts to reduce the complexity of dynamic models like (\ref{reduced2}) have introduced rules for selecting a subset of relevant EFMs, or grouping EFMs into families to be considered together (e.g. \cite{Song10,Song11,Vilkhovoy16}). An additional simplification is to approximate the composite EFM fluxes $r_k(\mathbf{m}_s)$ by Michaelis-Menten kinetics, such that
\begin{equation*}
r_k(\mathbf{m}_s) = V^{max}_k \prod_a \frac{m_a}{\kappa_{a,k} + m_a}
\end{equation*}            
where $V^{max}_k$, $\kappa_{a,k}$ are constants and the product includes all slow metabolite concentrations $m_a$ whose uptake fluxes are in the support of the $k$th EFM. In what follows it is assumed that such an approximation (although not necessarily the Michaelis-Menten one) has been provided for the functional form of the $r_k(\mathbf{m}_s)$ and that vectors representing EFMs have therefore been normalised to a common scale, such as total uptake carbon content. As described in Section \ref{sec:model}, choosing a common normalisation for the EFM representative vectors is essential when the $r_k$ are approximated in this way because then (\ref{reduced2}) is no longer invariant to $\mathbf{Z}^k$ re-scaling. The focus of this section is to understand the effect of further model reduction, by grouping EFMs into families, on resource allocation from the perspective of the maximum entropy control. Rules for composing EFM families are not the object of consideration here, but could involve, for example \cite{Song10,Song11,Vilkhovoy16}, grouping together all EFMs whose support contain the same uptake flux.

Partitioning of EFMs into $M$ families means partitioning indices $k=1,2,...,K$ into $M$ mutually disjoint subsets $F_J$ (of size $N_J$) $J=1,2,...,M$, and partitioning total resource into $M$ fractions $U_J$ such that
\begin{equation*}
\sum_{j \in F_J} u_j = U_J , \forall J =1,2,...,M : \quad \sum_{J=1}^M U_J = 1 .
\end{equation*}
Consequently, the EFM with index $j \in F_J$ is allocated a fraction $\tilde{u}_j = u_j/U_J$ of the resource $U_J$ available to the $J$th family. These values are collected in vectors $\mathbf{U} = (U_1,U_2,...,U_M)^T$ and $\mathbf{\tilde{u}}_J = (\tilde{u}_{F_{J,1}},\tilde{u}_{F_{J,2}},...,\tilde{u}_{F_{J,N_J}})^T$, where $F_{J,i}$ denotes the $i$th element of $F_J$. Representative vectors $\mathbf{\tilde{Z}}^J$ ($J=1,2,...,M$) are formed as weighted combinations of EFMs and used to express the dynamical system (\ref{reduced2}) in terms of EFM families. Several different weightings have previously been considered \cite{Song10,Song11,Vilkhovoy16}, but the entropy constraint identifies a particularly natural one to be
\begin{equation}
\label{weighting}
\tilde{r}_J(\mathbf{m}_{s})\mathbf{\tilde{Z}}^J = \frac{\sum_{j \in F_J} r_j(\mathbf{m}_{s}) \mathbf{Z}^ju_j }{\sum_{j \in F_J} u_j } = \sum_{j \in F_J} r_j(\mathbf{m}_{s}) \mathbf{Z}^j\tilde{u}_{F_{J,j}}
\end{equation}
where $\mathbf{\tilde{Z}}^J$ is defined given $\tilde{r}_J(\mathbf{m}_{s})$, $\mathbf{\tilde{u}}_J$, and the $r_j(\mathbf{m}_{s})$ ($j \in F_J$). To understand resource allocation in terms of this partitioning, observe the optimal control (\ref{control}) is obtained by maximisation of the objective functional
\begin{equation}
\label{objf}
\mathcal{F}(\mathbf{u}) =  \sum_{k=1}^K \mathcal{R}^k_{\Delta t}(\mathbf{m}_{s}) u_k + \sigma H(\mathbf{u})
\end{equation}
where the effective return-on-investment $\mathcal{R}^k_{\Delta t}(\mathbf{m}_{s})$ is defined in (\ref{eroi}). A classical result \cite{Jaynes57,Shore80} is that entropy satisfies the composition property 
\begin{equation}
\label{composition}
H(\mathbf{u}) = H(\mathbf{U}) + \sum_{J =1}^M U_J H(\mathbf{\tilde{u}}_J)
\end{equation}
where $H$ is defined as in (\ref{entropy}) on components of each respective vector. As shown in Appendix \ref{families}, combined with the weighting (\ref{weighting}) this implies $\mathcal{F}(\mathbf{u})$ can be expressed as 
\begin{equation}
\label{newobjf}
\mathcal{F}(\mathbf{u}) =  \sum_{J=1}^M \mathcal{F}_J(\mathbf{\tilde{u}}_J) U_J + \sigma H(\mathbf{U}) ,
\end{equation}
where $\mathcal{F}_J$ is the restriction of $\mathcal{F}$ to EFMs in the $J$th family. 

The full resource allocation problem can be viewed as a two-stage process involving an initial distribution of resource across EFM families, followed by further partitioning of the $U_J$ among their constituent EFMs \cite{Song10,Song11}. This procedure is best captured by maximising $\mathcal{F}(\mathbf{u})$ in two steps: first, the $\mathcal{F}_J(\mathbf{\tilde{u}}_J)$ are maximised for each family, providing maximum entropy controls for EFMs in terms of $\mathbf{\tilde{u}}_J$. Next, the resulting $\mathbf{\tilde{u}}_J$ are used for the EFM weightings (\ref{weighting}) and $\mathcal{F}(\mathbf{u})$ is maximised with respect to the $U_J$, yielding a maximum entropy control for EFM families. From a modelling perspective it is informative to consider the related objective functional 
\begin{equation}
\label{famobjf}
\mathcal{F}(\mathbf{U}) = \sum_{J=1}^M \mathcal{\tilde{R}}^J_{\Delta t}(\mathbf{m}_{s})U_J + \sigma H(\mathbf{U}) ,
\end{equation} 
where $\mathcal{\tilde{R}}^J_{\Delta t}(\mathbf{m}_{s})$ is the $J$th family's effective return-on-investment derived, as for individual EFMs in Section \ref{sec:control}, directly from system (\ref{reduced2}) expressed in terms of EFM families. Maximisation of $\mathcal{F}(\mathbf{U})$ provides the maximal entropy control to be used if the constituents of EFM families are not known, and in this case $\mathcal{\tilde{R}}^J_{\Delta t}(\mathbf{m}_{s})$ should be a suitable approximation of $\mathcal{F}_J(\mathbf{\tilde{u}}_J)$. For reasons described previously however, even when an enumeration of EFMs exists one might want to simplify the calculation of optimal controls by approximating $\mathcal{F}_J(\mathbf{\tilde{u}}_J)$ using EFMs in a manner consistent with maximisation. One way to do this is to set $r_k=x=1$ in the greedy effective return-on-investment (\ref{return}) and use the resulting (fixed) greedy maximum entropy controls 
\begin{equation}
\label{approxmaxent}
\tilde{u}_j = \frac{e^{\eta_j/\sigma}}{\sum_{l \in F_J} e^{\eta_l/\sigma}} \quad \forall j \in F_J , \quad J = 1,2,...,M
\end{equation}
with $\eta_j \equiv \mathbf{c}^T\mathbf{Z}^j$ to express the reduced dynamical system (\ref{reduced2}) in terms of EFM family vectors $\mathbf{\tilde{Z}}^J = \sum_{j \in F_j} \tilde{u}_j \mathbf{Z}^j$. The effective return-on-investment $\mathcal{\tilde{R}}^J_{\Delta t}(\mathbf{m}_{s})$ derived for the $J$th family then takes the same form as (\ref{eroi}), but with $B^k$ replaced by 
\begin{equation}
\label{BJ}
\tilde{B}^J = x \tilde{r}_J(\mathbf{m}_{ex}) \begin{pmatrix}  \mathbf{S}_{ex}  \\ \mathbf{c}^T  \end{pmatrix}  \mathbf{\tilde{Z}}^J .
\end{equation}
Maximisation of the objective functional $\mathcal{F}(\mathbf{U})$ in (\ref{famobjf}) therefore provides the maximum entropy control for optimal allocation of resource among EFM families represented by these $ \mathbf{\tilde{Z}}^J$. This approach is analogous to the EFM lumping proposed by Song and Ramkrishna \cite{Song10,Song11}, where it is of relevance to note that their choice of EFM weighting also depends on a parameter $n_v$ that modulates spread across EFMs within a family, thus playing the equivalent of $\sigma$ in the maximal entropy control. In fact, their choice of fixed weighting $u_j = \eta_j^{n_v}$ is related to the fixed maximum entropy weighting $u_j \propto e^{\eta_j/\sigma}$ in the same way that maximum entropy relates to fuzzy clustering \cite{Karayiannis94}. An alternative method for approximating $\mathbf{\tilde{Z}}^J$, related to \cite{Vilkhovoy16} and the maximum entropy control in the $\sigma \to 0$ limit, is to select the vector $\mathbf{Z}^j$ ($j\in F_J$) representing the EFM with the largest return-on-investment (i.e., the FBA/Bang-Bang solution) in the $J$th family.

Although EFM families and various EFM weightings have been described previously \cite{Song10,Song11,Vilkhovoy16}, the above discussion implicates their unification under the maximal entropy control framework. Stated in terms of a modelling endeavour to best approximate $\mathcal{F}_J(\mathbf{\tilde{u}}_J)$, choices for individual EFM return-on-investments and EFM family representatives can be derived from first principles. These approximations may involve any form of dynamic model reduction or simplification that respects the basic maximum entropy control laws for partitioning resources among EFMs within a family. Moreover, subsequent partitioning of resource among EFM families has always previously involved the original cybernetic control laws of Young and Ramkrishna \cite{Young07}, whereas for consistency the maximum entropy control should once more be used to determine the $U_J$. Doing so makes it possible to recursively apply this form of model reduction when establishing the appropriate number of EFM families; the recursive nature of the maximum entropy control framework is succinctly captured by the objective functional (\ref{newobjf}).             

\begin{exmp}
\label{example4}
Consider the metabolic network (Figure \ref{fig:2a}) from Example \ref{example3}, but now with $C$ interpreted as the concentration of a {\em fast} intracellular metabolite. The QSSA on $C$ imposes the additional algebraic constraint $v_5 - v_6 = 0$, which results in the collapse of one EFM (represented by vector $\mathbf{Z}^6$ in Example \ref{example3}). A complete set of five EFMs for this network is therefore provided by the representative vectors
\begin{equation*}
\mathbf{Z}^1 =  \frac{1}{6} \begin{pmatrix} 1 \\ 1 \\ 2 \\ 0 \\ 0 \\ 0 \\ 0 \end{pmatrix} , \quad \mathbf{Z}^2 =  \frac{1}{3} \begin{pmatrix}0 \\ 0 \\ 0 \\ 1 \\ 1 \\ 0 \\ 0  \end{pmatrix} , \quad \mathbf{Z}^3 =   \frac{1}{6} \begin{pmatrix} 1 \\ 1 \\ 0 \\ 2 \\ 0 \\ 0 \\ 0 \end{pmatrix} , \quad  \mathbf{Z}^4 =  \frac{1}{6} \begin{pmatrix} 1 \\ 1 \\ 2 \\ 0 \\ 0 \\ 1 \\ 1  \end{pmatrix} , \quad \mathbf{Z}^5 =   \frac{1}{6} \begin{pmatrix} 1 \\ 1 \\ 0 \\ 2 \\ 0 \\ 1 \\ 1 \end{pmatrix} ,
\end{equation*}
each normalised by their uptake carbon content. Partitioning of these EFMs into families on the basis of their shared products and substrates means $F_1 = \{1,4\}$, $F_2 = \{2\}$, and $F_3 = \{3,5\}$, which gives rise to the three EFM families represented in Figure \ref{fig:3}.
\begin{figure}
    \caption{Partitioning of the five EFMs from the metabolic network in Example \ref{example3} with $C$ interpreted as a fast metabolite. The five EFMs are represented by vectors $\mathbf{Z}^k$ (dashed lines) and each of the three EFM families is represented by an individual vector $\mathbf{\tilde{Z}}^J$ (bold lines). In this case $\mathbf{\tilde{Z}}^2 = \mathbf{Z}^2$, while $\mathbf{\tilde{Z}}^1$ and $\mathbf{\tilde{Z}}^3$ are formed from linear combinations of the pairs $(\mathbf{Z}^1,\mathbf{Z}^4)$ and $(\mathbf{Z}^3,\mathbf{Z}^5)$, respectively, as described in Example \ref{example4}.} \label{fig:3}

     \begin{subfigure}[t]{0.65\textwidth}
     \centering
          \caption{Representation of the first EFM family by the vector $\mathbf{\tilde{Z}}^1$.} \label{fig:3a}
     \includegraphics[width=\linewidth]{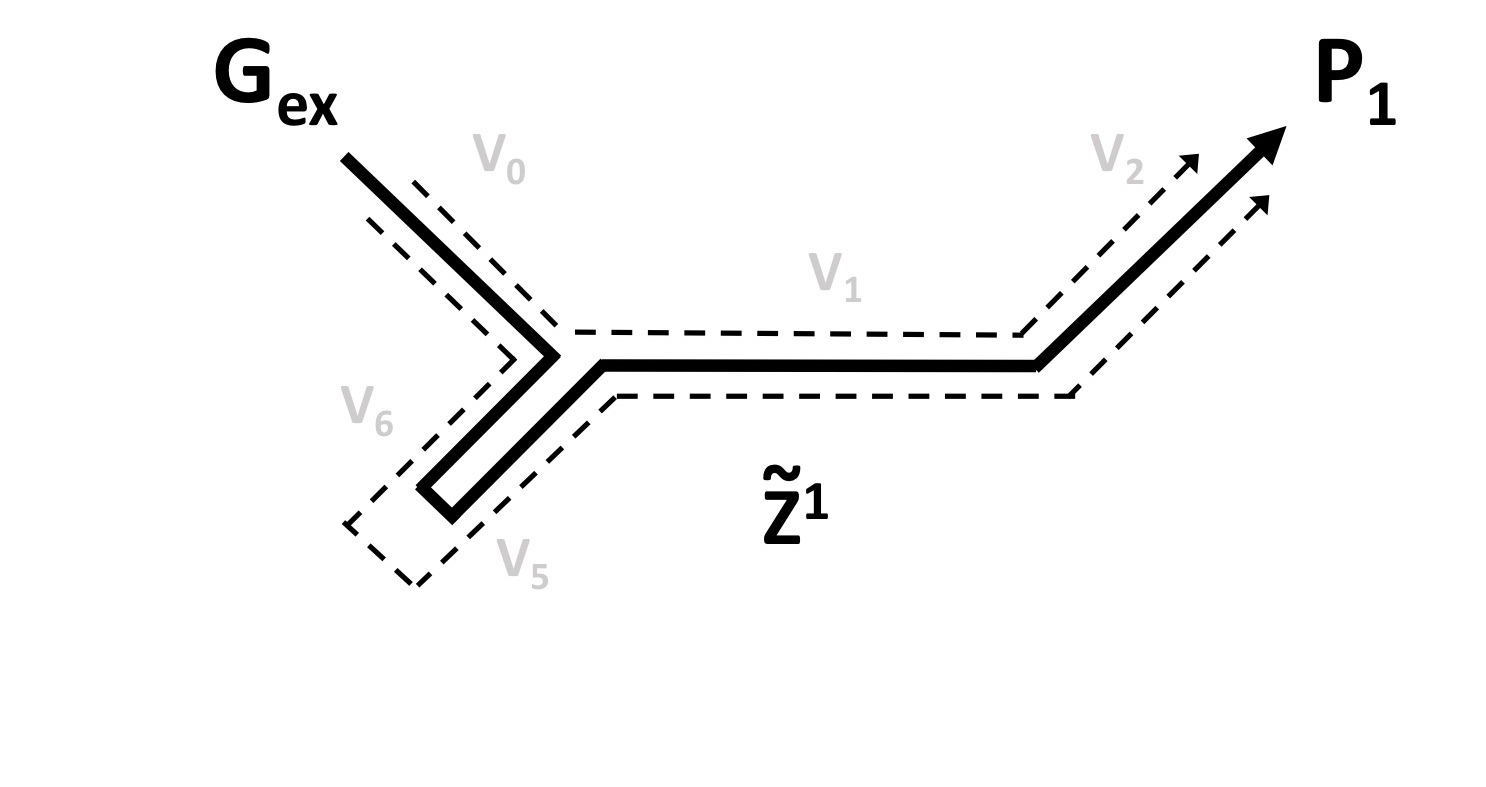} 
    \end{subfigure}
    \hfill
    \begin{subfigure}[t]{0.275\textwidth}
        \centering
                \caption{Representation of the second EFM family by the vector $\mathbf{\tilde{Z}}^2$.} \label{fig:3b}
        \includegraphics[width=\linewidth]{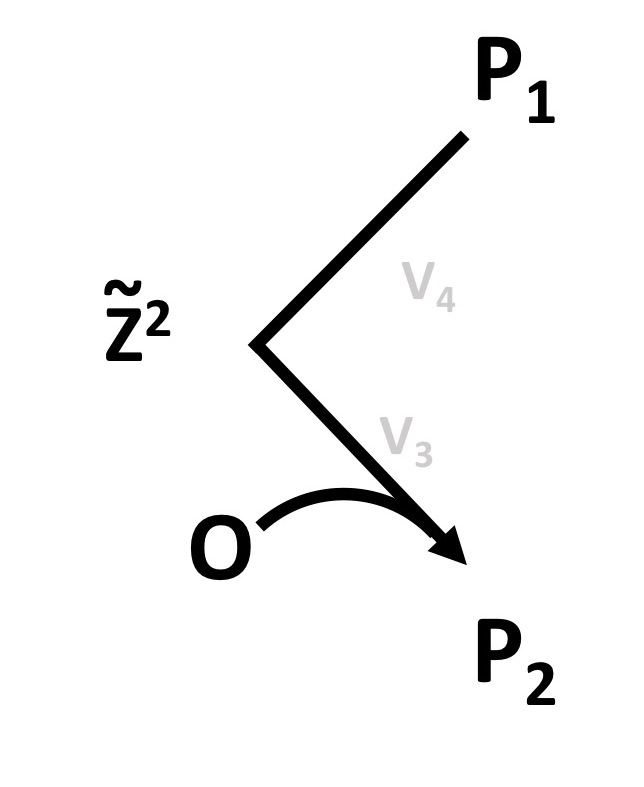} 
    \end{subfigure}
        \vspace{1cm}
     \centering
    \begin{subfigure}[t]{0.65\textwidth}
            \caption{Representation of the third EFM family by the vector $\mathbf{\tilde{Z}}^3$.} \label{fig:3c}
    \centering
        \includegraphics[width=\linewidth]{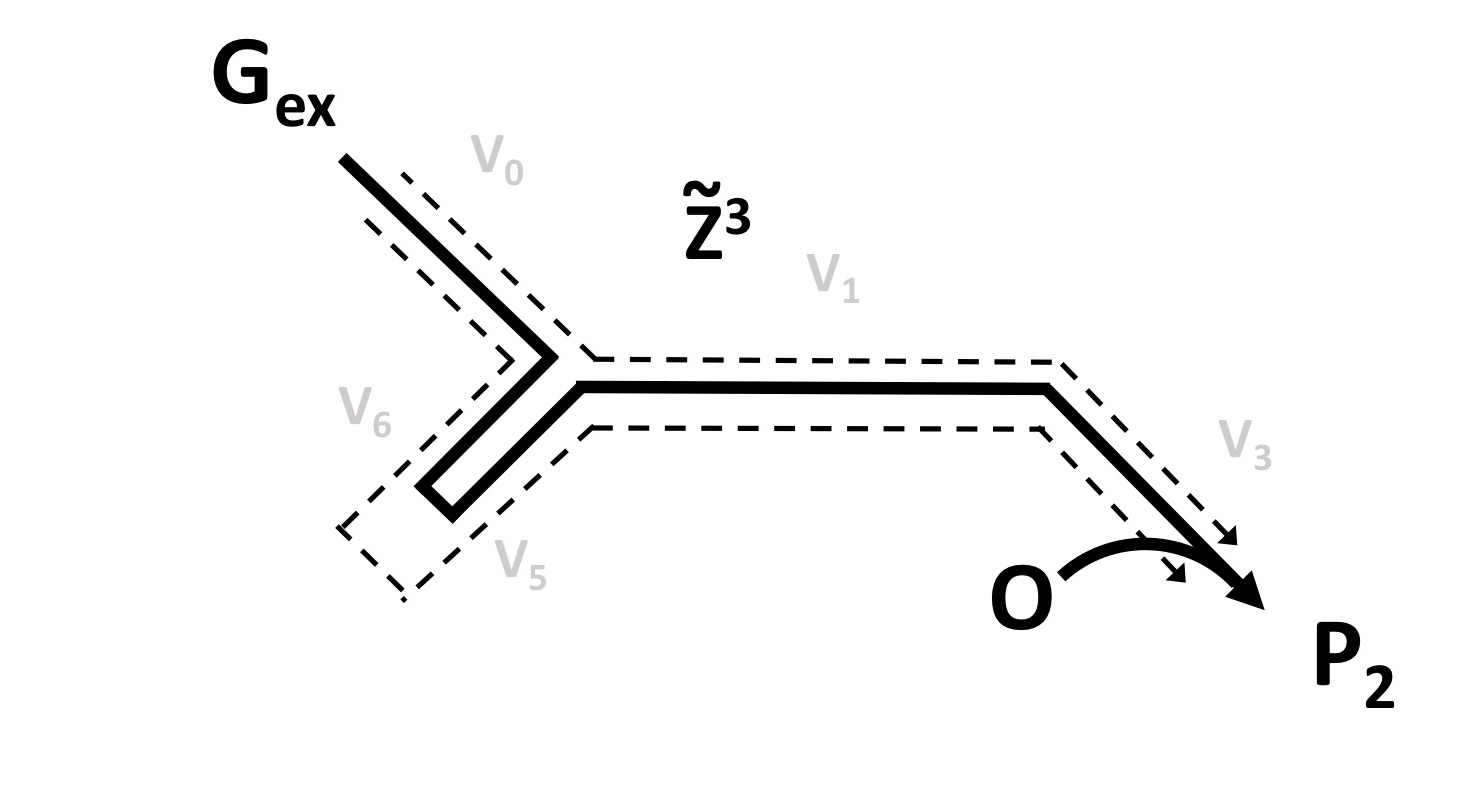} 
    \end{subfigure}
        
\end{figure}
Setting $r_k=x=1$ ($k =1,2,3,4,5$) in (\ref{return}) yields the fixed return-on-investments
\begin{equation*}
\eta_1 = \eta_4 = \frac{1}{6}c_1, \quad \eta_2 = \frac{1}{3}c_3,   \quad \eta_3 = \eta_5 = \frac{1}{6}(c_1 + 2c_3) 
\end{equation*}
and in this case the fixed greedy maximum entropy controls (\ref{approxmaxent}) for each family are independent of $\sigma$:
\begin{equation*}
\tilde{u}_1 = \tilde{u}_4 = \frac{1}{2}, \quad \tilde{u}_2 = 1,  \quad  \tilde{u}_3 = \tilde{u}_5 = \frac{1}{2} .
\end{equation*}
This results in the three EFM family representative vectors 
\begin{equation*}
\mathbf{\tilde{Z}}^1 = \frac{1}{2}( \mathbf{Z}^1 +   \mathbf{Z}^4), \quad \mathbf{\tilde{Z}}^2 = \mathbf{Z}^2, \quad  \mathbf{\tilde{Z}}^3 =  \frac{1}{2} ( \mathbf{Z}^3 +  \mathbf{Z}^5)
\end{equation*}
and using a Michaelis-Menten approximation for the $\tilde{r}_J$ one has
\begin{eqnarray*}
\tilde{r}_1(G_{ex}) &=& \tilde{V}^{max}_1 \frac{G_{ex}}{\tilde{\kappa}_{G,1} + G_{ex}}\\
\tilde{r}_2(O,P_1) &=& \tilde{V}^{max}_2 \frac{O}{\tilde{\kappa}_{O,2} + O} \frac{P_1}{\tilde{\kappa}_{P,2} + P_1}  \\
\tilde{r}_3(G_{ex},O) &=& \tilde{V}^{max}_3 \frac{G_{ex}}{\tilde{\kappa}_{G,3} + G_{ex}} \frac{O}{\tilde{\kappa}_{O,3} + O} .
\end{eqnarray*}
Substituting for these expressions in (\ref{BJ}) provides $\tilde{B}^{J}$ to be used for calculating the effective return-on-investment of the $J$th EFM family and maximum entropy controls $U_1,U_2,U_3$. Note that in this example, independence of $\sigma$ in the fixed greedy maximum entropy controls $\mathbf{\tilde{u}}_1$, $\mathbf{\tilde{u}}_2$, $\mathbf{\tilde{u}}_3$ implies they are equivalent to a set of FBA/Bang-Bang policies. Alternatively, one could assume some cost is associated with one of the reactions $v_5$ or $v_6$ (e.g., $c_5 < 0$), in which case $\eta_1 > \eta_4$ and $\eta_1 > \eta_5$ so that this equivalence would only hold in the $\sigma \to 0$ limit where $\mathbf{\tilde{Z}}^1 = \mathbf{Z}^1$ and $\mathbf{\tilde{Z}}^3 = \mathbf{Z}^3$. In both cases the reduced dynamic model can be mapped on to that described in Example \ref{example1}.       

\qed
\end{exmp}

\section{Application to yeast metabolism}
\label{sec:appplication}
Here a dynamic model of resource allocation is introduced for central carbon metabolism in a single-celled eukaryotic organism, yeast, which builds upon the concepts and examples presented in previous sections. The metabolic network in Figure \ref{fig:2a} describes the participating reactions, and product 1 is ethanol (concentration $E$), with $C$ the lumped concentration of storage carbohydrates glycogen and trehalose modelled in this case (contrast with Example \ref{example3}) as an extracellular term that combines with total catalytic biomass $x$ to represent total biomass $x +C$ in the system. Note that without storage carbohydrates, the simplified network (Figure \ref{fig:1a}) may be viewed as the reduction of a much larger network composed of glycolysis, the pentose-phosphate pathway, citric acid cycle, glyoxylate shunt, and oxidative phosphorylation, using EFM families exactly as was suggested by Song and Ramkrishna \cite{Song10}. Introduction of the two reactions with fluxes $v_5,v_6$ follows this same principle by grouping all reserve carbohydrate pathways into a single representative family. Precise details of EFM family reduction have been overlooked but, because of its recursive nature as described in the previous section, the relevant EFM families can be represented by the six vectors $\mathbf{Z}^k$ from Example \ref{example2} normalised by their uptake carbon content. The full dynamical system expressed in terms of these (with the subscript dropped from $G_{ex}$) is
\begin{eqnarray*}
\frac{dG}{dt} &=& D(G_0 - G) -\frac{1}{2}[ r_1u_1 + r_3u_3 + r_6u_6] x\\
\frac{dO}{dt} &=&   kLa(O^* - O) -[r_2u_2 + r_3u_3 + r_5u_5] x  \\
\frac{dE}{dt} &=& -DE+[r_1u_1 -  r_2u_2 + r_4u_4]x   \\
\frac{dC}{dt} &=& -DC+ \frac{1}{2}[r_6u_6 -  r_4u_4 - r_5u_5]x   \\
\frac{dx}{dt} &=& [\frac{1}{2}c_1(r_1u_1 + r_4u_4)  + c_3  r_2u_2 + \frac{1}{2}(c_1 + 2c_3) ( r_3u_3 + r_5u_5) - D]x
\end{eqnarray*}
where a dilution rate $D$ and volumetric mass transfer coefficient $k_La$ have been introduced for simulation of cultures with inflow glucose concentration $G_0$ and dissolved oxygen solubility limit $O^*$. 

The $r_k$ are approximated by Michaelis-Menten kinetics according to the metabolites whose uptake fluxes are in the support of each $\mathbf{Z}^k$, such that
\begin{equation*}
r_1(G) = V^{max}_1 \frac{G}{K_1 + G}, \quad r_2(O,E) = V^{max}_2 \frac{E}{K_2 + E}\frac{O}{K_{O,2} + O}
\end{equation*}

\begin{equation*}
r_3(G,O) = V^{max}_3 \frac{G}{K_3 + G}\frac{O}{K_{O,3} + O}, \quad r_4(C/x) = V^{max}_4 \frac{C/x}{K_4 + C/x}
\end{equation*}

\begin{equation*}
r_5(O,C/x) = V^{max}_5 \frac{C/x}{K_5 + C/x}\frac{O}{K_{O,5} + O}, \quad r_6(G) = V^{max}_6 \frac{G}{K_6 + G} .
\end{equation*}
Observe that composite fluxes $r_4,r_5$ depend on the fractional concentration $C/x$ of storage carbohydrate. The zeroth-order return-on-investments are
\begin{equation*}
R^1_0(G) =  \frac{1}{2}c_1 r_1(G), \quad R^2_0(O,E) = c_3 r_2(O,E) , 
\end{equation*}  

\begin{equation*}
R^3_0(G,O) = \frac{1}{2}(c_1 + 2c_3) r_3(G,O) , \quad R^4_0(C/x) = \frac{1}{2}c_1 r_4(C/x) , 
\end{equation*}

\begin{equation*}
R^5_0(O,C/x) = \frac{1}{2}(c_1 + 2c_3) r_5(O,C/x) , \quad R^6_0 = 0 
\end{equation*} 
and the greedy maximum entropy control law for determining each $u_k$ is then
\begin{equation*}
u_k = \frac{\exp(xR^k_0(\mathbf{m},x)/\sigma)}{\sum_{l=1}^6 \exp(x R^l_0(\mathbf{m},x)/\sigma)}
\end{equation*} 
with $\mathbf{m} = (G,O,E,C)^T$ and $\sigma > 0 $. Numerical simulations of this system were performed using custom-built software based on the SUNDIALS solvers \cite{Hindmarsh05} and the fixed parameter values listed in Table \ref{tab:table1}. Note that parameter values have been chosen generically in the sense that, although biologically realistic values have been used, there has been no attempt to fit these to experimental data or perform bifurcation analysis. The results of simulations should therefore be treated as qualitative and absolute concentration measurements used for quantitative predictions only after using biological knowledge or data to refine parameter values, which will almost certainly increase overall predictive power of the model. In particular, it is known that modifying parameters $V_1^{max}, V_2^{max}, V_3^{max}$ or introducing a cost of each pathway (i.e., weights for $u_1,u_2,u_3$ in the summation constraint from (\ref{linprob})) generates a static model of resource allocation for the network in Example \ref{example1} where either pure oxidation, pure fermentation, or a mixture of oxidation and fermentation, are optimal strategies for maximal ATP production (see \cite{Moller18} and references therein). This result is based on the proposal that overflow metabolism is a consequence of lower yield- higher rate pathways being preferred when certain environmental conditions are combined with the constraints of resource allocation \cite{Basan15}. By contrast, simulations based on the parameter values in Table \ref{tab:table1} are only intended to capture isolated effects of the maximum entropy control law without imposing any additional biological information.          

\begin{table}[h!]
  \begin{center}
    \caption{Values for parameters used in all simulations. Initial conditions and values for $D$, $k_La$, $G_0$, and $\sigma$ are reported in the legend of Figure \ref{fig:4}.}
    \label{tab:table1}
    \begin{tabular}{lr}
      \textbf{Parameters} & \textbf{Value} \\
      \hline
      $V_1^{max},V_2^{max},V_3^{max},V_4^{max},V_5^{max}$ & $1.0$ $\mbox{h}^{-1}$ \\
      $V_6^{max}$ & $2.5$ $\mbox{h}^{-1}$ \\
      $c_1$ & $0.02$ $\mbox{g}\cdot \mbox{g}^{-1}$ \\
      $c_3$ & $0.34$ $\mbox{g} \cdot \mbox{g}^{-1}$ \\
      $K_1,K_2,K_3,K_6$ & $0.01$ $\mbox{g} \cdot \mbox{L}^{-1}$ \\
      $K_4,K_5$ & $0.01$ $\mbox{g}  \cdot \mbox{L}^{-1} \cdot \mbox{g}^{-1} \cdot \mbox{L} $ \\
      $K_{O,2},K_{O,3},K_{O,5}$ & $0.001$ $\mbox{g} \cdot \mbox{L}^{-1}$ \\
      $O^*$ & $0.015$ $\mbox{g} \cdot \mbox{L}^{-1}$ \\
    \end{tabular}
  \end{center}
\end{table}

\begin{figure}
    \caption{Results of numerical simulation for various values of $\sigma$ using the parameters reported in Table \ref{tab:table1} with: $D=0.0$, $k_La = 30.0$ $\mbox{h}^{-1}$, $G_0 = 0$ $\mbox{g} \cdot \mbox{L}^{-1}$ and initial conditions $x(0)=0.1$, $G(0) = 10.0$, $O(0) = 0.0001$, $E(0) = 0.0$ $\mbox{g} \cdot \mbox{L}^{-1}$ and $C(0) = 0.0$ $\mbox{g} \cdot \mbox{g}^{-1}\cdot  \mbox{L}^{-1} $ for batch culture; and $D = 0.1 - 0.135$ $\mbox{h}^{-1}$, $G_0 = 10.0$ $\mbox{g} \cdot \mbox{L}^{-1}$, $k_La = 150.0$ $\mbox{h}^{-1}$ and initial conditions $x(0)=0.1$, $G(0) = 0.04$, $O(0) = 0.001$, $E(0) = 0.0$ $\mbox{g} \cdot \mbox{L}^{-1}$ and $C(0) = 0.0$ $\mbox{g} \cdot \mbox{g}^{-1}\cdot  \mbox{L}^{-1} $ for continuous culture (at steady state).} \label{fig:4}

     \begin{subfigure}[t]{0.475\textwidth}
     \centering
          \caption{Batch culture with $\sigma = 0.001$} \label{fig:4a}
     \includegraphics[width=\linewidth]{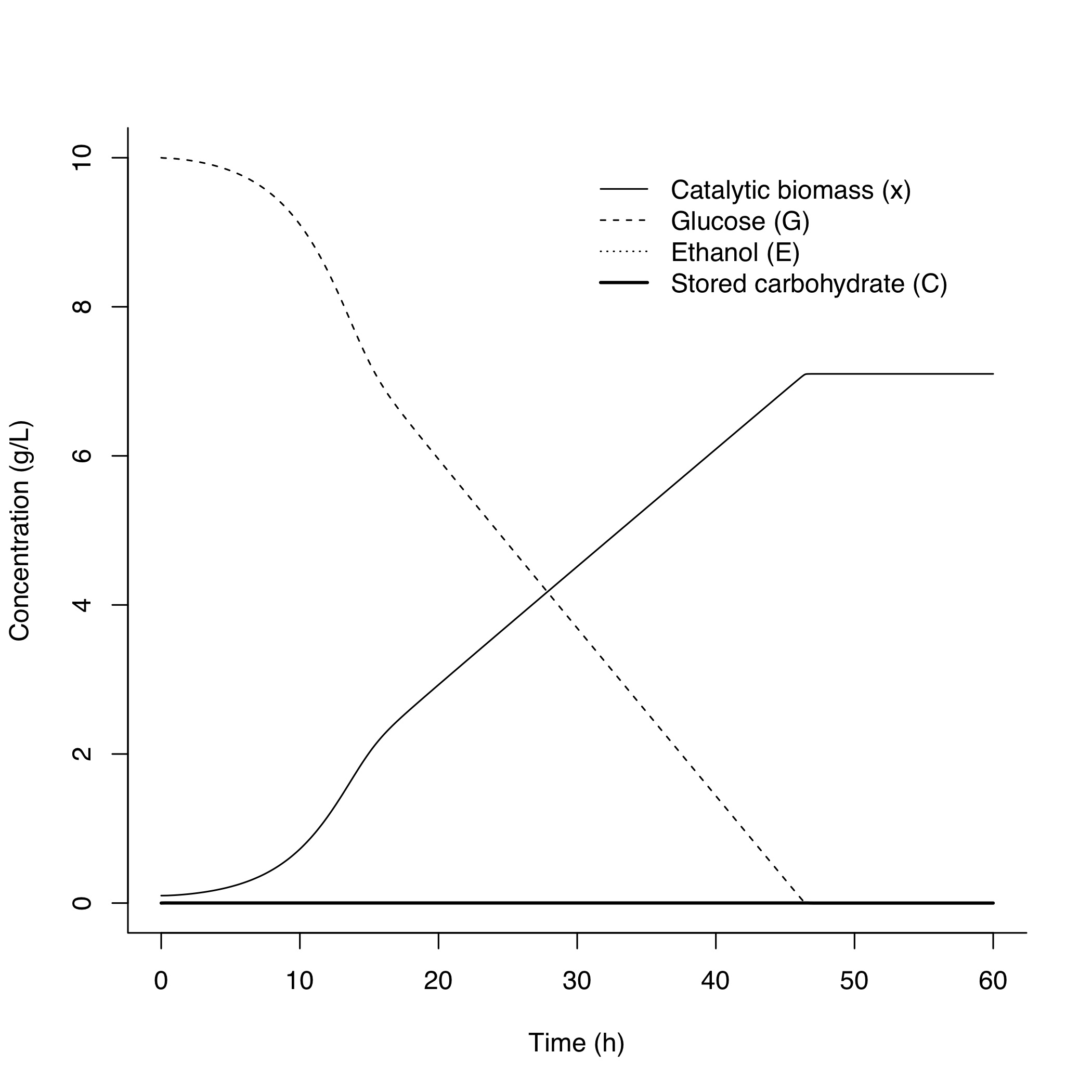} 
    \end{subfigure}
    \hfill
    \begin{subfigure}[t]{0.475\textwidth}
        \centering
                \caption{Batch culture with $\sigma = 0.1$} \label{fig:4b}
        \includegraphics[width=\linewidth]{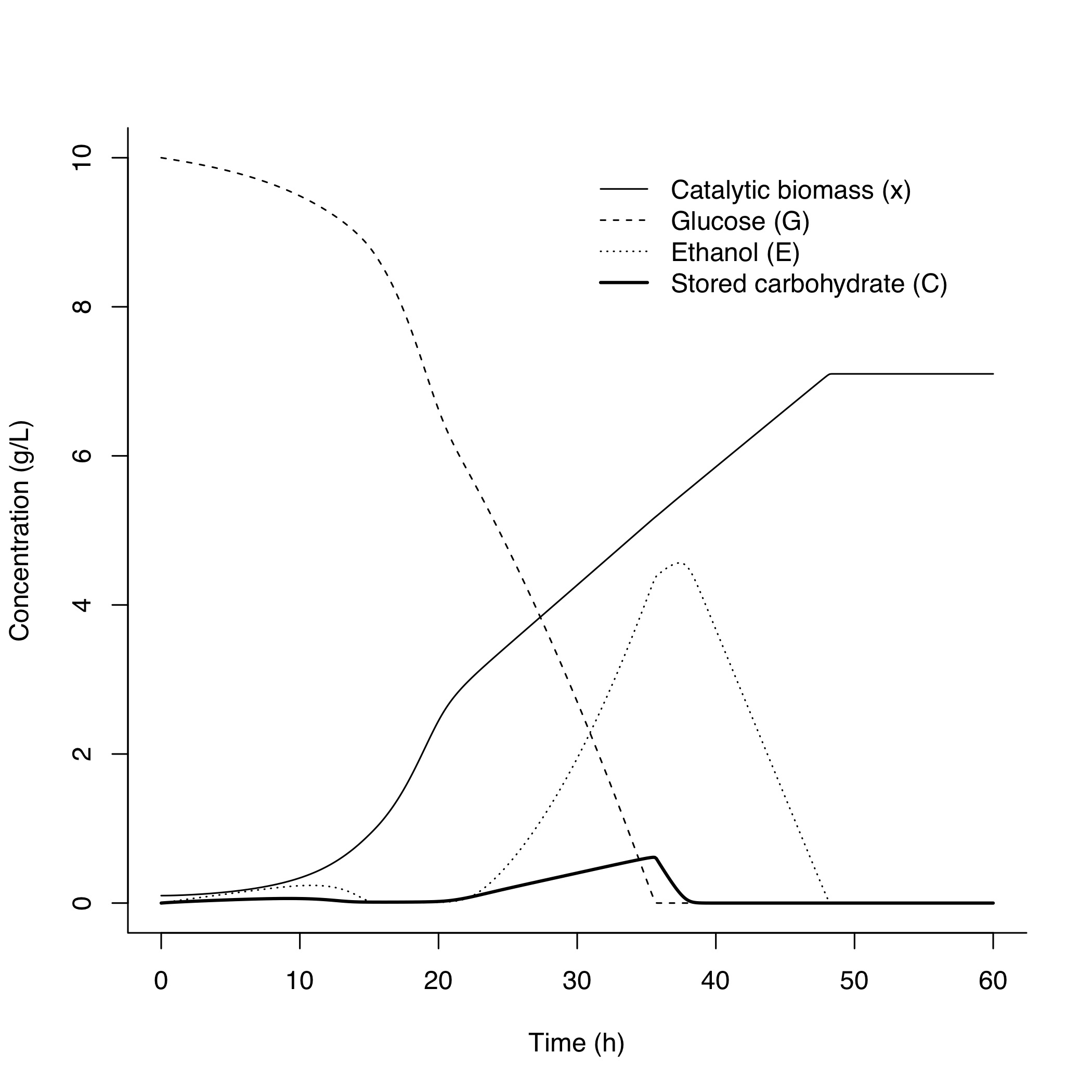} 
    \end{subfigure}
        \vspace{1cm}
     \centering
    \begin{subfigure}[t]{0.475\textwidth}
     \centering
          \caption{Batch culture with $\sigma = 1.0$} \label{fig:4c}
     \includegraphics[width=\linewidth]{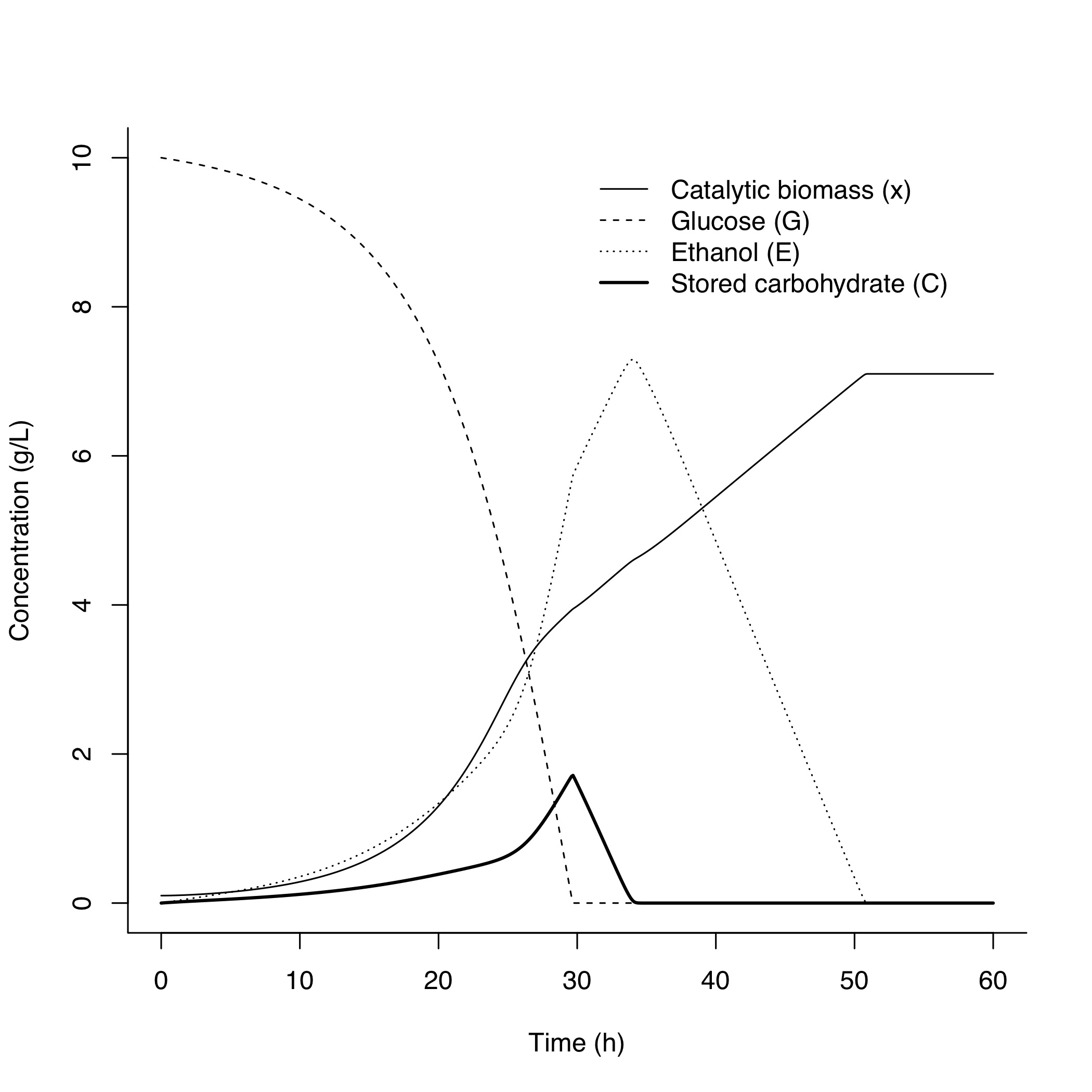} 
    \end{subfigure}
    \hfill
    \begin{subfigure}[t]{0.475\textwidth}
            \caption{Continuous culture with $\sigma = 1.0$} \label{fig:4d}
    \centering
        \includegraphics[width=\linewidth]{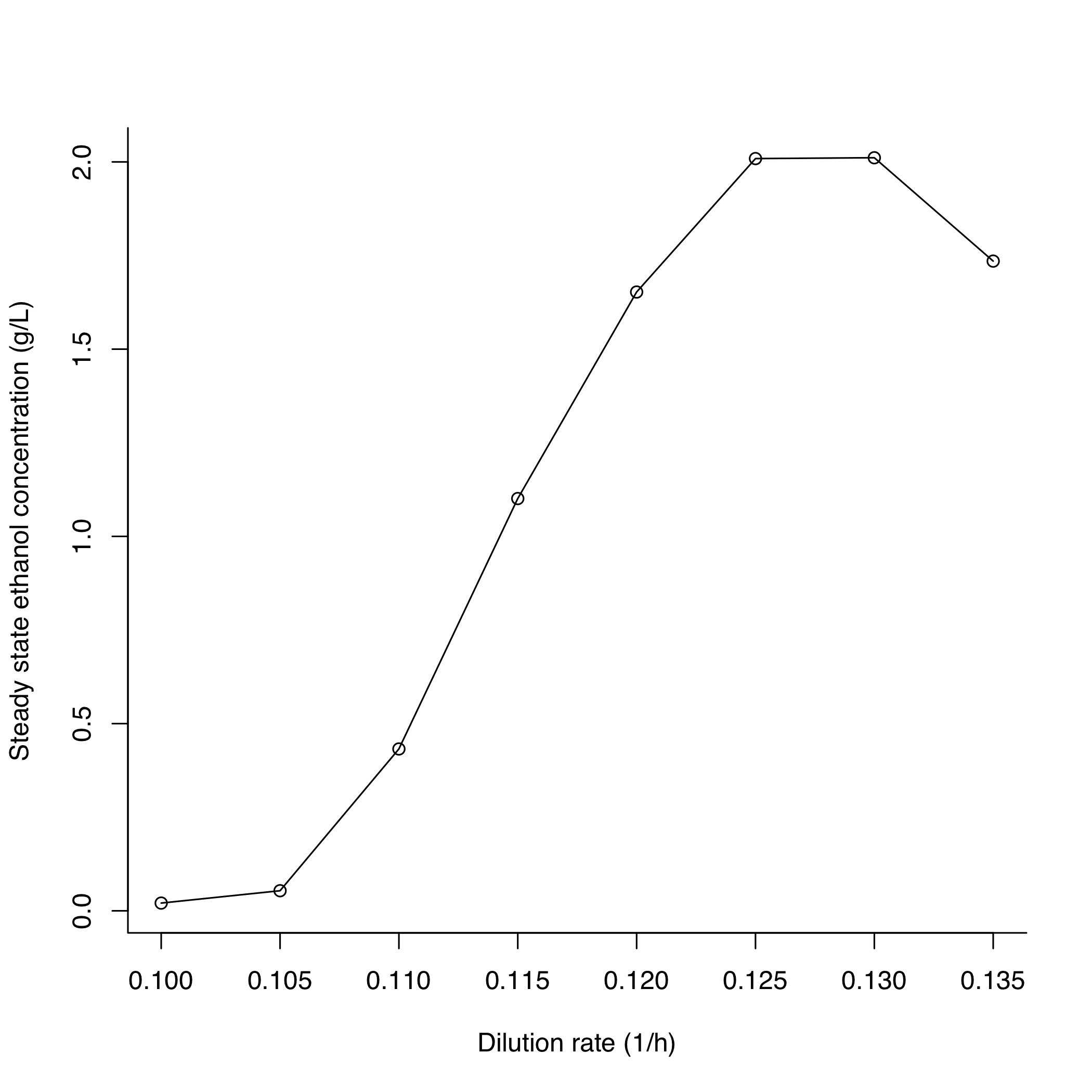} 
    \end{subfigure}
        
\end{figure}

Figures \ref{fig:4a}-\ref{fig:4c} show batch culture simulations for increasing values of $\sigma$.
When $\sigma$ is small (Figure \ref{fig:4a}), there is relatively little accumulation of ethanol or storage carbohydrate over the course of simulation because, for the given parameter values at the provided oxygen concentrations, the glucose oxidation pathway (EFM represented by $\mathbf{Z}^3$) continues to receive the majority of resource until glucose is depleted and growth halted. The fraction of resource allocated to glucose fermentation (EFM represented by $\mathbf{Z}^1$) only becomes comparable towards the end of the simulation. When $\sigma$ is increased to an intermediate value however (Figure \ref{fig:4b}), there is a large accumulation of extracellular ethanol, which in turn feeds in to the ethanol oxidation pathway (EFM represented by $\mathbf{Z}^2$) that receives the majority of resource during the latter half of the simulation. This transition from glucose metabolism to ethanol oxidation qualitatively captures the Crabtree effect observed in batch cultures of baker's or brewer's yeast {\em Saccharomyces cerevisiae} \cite{Pfeiffer14}, but note that this effect could have been reproduced at lower values of $\sigma$ should $V_1^{max} >> V_3^{max}$, $K_1 >> K_3$, or the cost of the oxidative pathways (weights for $u_2,u_3$ in (\ref{linprob})) be increased in analogy with the model of overflow metabolism presented in \cite{Moller18}. Instead, for the parameter values reported here, ethanol accumulation and the subsequent transition between substrates is solely attributable to the maximum entropy control law, which allocates non-zero fractions of resource to EFMs with lower return-on-investment. This behaviour is indicative of a bet-hedging component of the Crabtree effect, possibly related to that observed experimentally in the case of the diauxic shift \cite{Solopova14}. When $\sigma$ is increased further (Figure \ref{fig:4c}), there is a considerable accumulation of storage carbohydrate that leads to the pathways involved in its consumption (EFMs represented by $\mathbf{Z}^4$ and $\mathbf{Z}^5$) receiving the majority of resource for a brief window of time at the point of transition from glucose to ethanol. Accumulation of storage carbohydrate occurs because a greater fraction of resource is allocated to the EFM represented by $\mathbf{Z}^6$, even though this pathway does not contribute directly to growth. Closer inspection reveals a sharp increase in the rate of storage carbohydrate accumulation immediately prior to the onset of the transition from glucose to ethanol and its subsequent utilisation, which agrees very well with the observed dynamics of glycogen and trehalose metabolism in {\em S. cerevisiae} batch culture: these metabolites rapidly accumulate near the end of the growth phase on glucose and are quickly consumed when consumption of ethanol begins \cite{Lillie80,Francios01,Panek62,Panek63,Parrou99}. Figure \ref{fig:4d} plots steady state concentrations of extracellular ethanol against dilution rate for simulations of the model in continuous culture with $\sigma = 1.0$. As a general trend, increasing dilution rate leads to an increase in extracellular ethanol concentrations before reaching a maximum value and decreasing prior to wash out (when dilution rate exceeds growth rate). In the future, these continuous culture simulations could be compared with data reported in \cite{Hoek98} for further refinement of parameter values.       

For comparison with other models of yeast metabolism at this level of complexity, two closely related examples are the model derived using EFM lumping by Song and Ramkrishna in \cite{Song10} and the model by Jones and Kompala \cite{Jones99}. Both are based on cybernetic laws originating from the greedy control of Young and Ramkrishna \cite{Young07}, but unlike the maximum entropy model presented here have been enlarged to include auxiliary variables (called `cybernetic enzymes') responsible for conferring the regulatory effects of control. Cybernetic enzymes may confer additional robustness to dynamic models, but also imply the existence of extra parameter values that are especially difficult to determine experimentally because they do not correspond to biological reality. One could argue that the EFM control variables $u_k$ do not correspond directly to biological entities either, but in this case they come without the overhead of additional uninterpretable parameters. Although based on EFM families, the model of Song and Ramkrishna \cite{Song10} does not include the dynamics of storage carbohydrates. In any case, their cybernetic control law would negate the allocation of resource to any EFMs involved in storage carbohydrate formation unless additional assumptions were imposed upon the model. Parameters used in their simulations were only partially informed by experimental data. On the other hand, the model of Jones and Kompala \cite{Jones99} does include a term for storage carbohydrates, but the coupling of $C$ to other variables in the model (structurally equivalent to that of Song and Ramkrishna) is based entirely on empirical observations \cite{Lillie80,Francios01,Panek62,Panek63,Parrou99} for the dynamics of carbohydrate accumulation and utilisation. Conversely, the description of this phenomenon presented here relies on nothing more than the maximum entropy principle. As a final note, parameter values in the Jones and Kompala model were obtained by direct fit to experimental data. Such an approach could be expected to improve the quantitate behaviour of the maximum entropy model and the lumped cybernetic model in \cite{Song10}, both to be viewed as qualitatively predictive in nature. 

\section{Conclusion}
Using the maximum entropy principle to extend the optimal control framework of Young and Ramkrishna \cite{Young07} provides a dynamic theory of metabolic resource allocation in the face of environmental uncertainty. This concept both generalises and unifies prior approaches by establishing a smooth interpolation between DFBA \cite{Mahadevan02} at one extreme, and dynamic metabolic models without regulation \cite{Provost04,Provost06} at the other. In contrast to alternative optimal control laws, no assumption other than instantaneous maximisation of total catalytic biomass based on the maximum entropy principle is required to explain activation of pathways not contributing directly to growth rate, such as the formation of reserve compounds that have previously been explicitly included as an integral component of total biomass. From an evolutionary perspective this is a particularly appealing explanation for the observed accumulation of reserve compounds in growth-limiting conditions, because selection for maximal rates of self-replication extends beyond cellular biology to the RNA world \cite{Orgel04}. However, it likely that this form of bet-hedging constitutes just one component of the biological mechanism that governs dynamic regulation of metabolism in fluctuating environmental conditions.   

Application of the dynamic maximum entropy framework to a simplified model of yeast metabolism has shown that the theory successfully reproduces some observed behaviour of cell populations in batch and continuous culture. This reduced model can almost surely be improved by including additional biological knowledge on the nature of overflow metabolism in yeast \cite{Moller18,Basan15,Pfeiffer14}, but also serves to illustrate contributions that come from bet-hedging alone. When considering cell populations, there are at least two (not mutually exclusive) theoretical interpretations for such bet-hedging mechanisms \cite{Hansen01,Sims03}: that the maximum entropy distribution of resources is a result of heterogeneity in the regulatory FBA/Bang-Bang strategies of individuals \cite{Ackermann15,Martins15,DeMartino17,Fernandez19}, or is chosen by each cell as the optimal strategy for dealing with uncertainty in the environment \cite{Solopova14,Granados17}. It is important to note that, if thermodynamic constraints are imposed, then only the FBA/Bang-Bang policy will be guaranteed to yield a thermodynamically-consistent, single-cell resource allocation strategy provided the set of EFMs is restricted to those that are thermodynamically-feasible \cite{Jol12}. This observation therefore promotes the interpretation that each individual in the population adopts an FBA/Bang-Bang policy \cite{Giordano16}, and that the relative fraction of total resource allocated to an EFM reflects the fraction of individuals within the population investing exclusively in that metabolic pathway. Due to the overall scaling effect of total catalytic biomass concentration $x$ in the effective return-on-investment, the concentration of resource on the EFM with largest return-on-investment will increase as $x$ does also, and therefore the maximum entropy resource allocation strategy of the entire population also approaches the FBA/Bang-Bang policy as the population grows in size. This leads to further concentration of resource on the pathway with greatest contribution to growth, whereas, when the population is growth-limited, resource allocation based on the temporal maximum entropy control is increased to pathways with higher total metabolite yield.   

Entropy is the uniquely-defined continuous function of a discrete probability distribution, monotonically increasing in the number of states when each occurs with equal probability, which respects the composition law of fractional partitioning \cite{Jaynes57,Shore80}. For applications to dynamic resource allocation this naturally identifies the maximum entropy distribution as the appropriate control law for performing model reduction based on EFM families, where several alternative EFM weightings have been introduced previously \cite{Song10,Song11,Vilkhovoy16}. The suggested EFM weighting based on the maximum entropy control law generalises those considered in prior work. Using EFM families becomes particularly important for larger metabolic networks where an explosion in the number of EFMs \cite{Klamt02} makes direct model parameterisation infeasible; however, the maximum entropy framework provides a consistent methodology for recursive model reduction. It is of worth pointing out that DeVilbiss and Ramkrishna have recently proposed an information theory-based model selection scheme \cite{DeVilbiss17}, using as a test case metabolic models expressed in terms of EFMs and the control laws of Young and Ramkrishna. Models based on EFM families were shown to provide the most succinct description of steady state fluxes as measured by information-theoretic criteria, and it would therefore be intriguing to explore how these criteria synergise with the information-theoretic concept of the maximum entropy control.                                           

\section*{Acknowledgements}
This work benefited from advice from JD Young, W Liebermeister, and P Dixit on metabolic modelling, and discussions with JS O'Neill and HC Causton on yeast metabolism. Thanks are extended to JD Young for also sharing their PhD Thesis, and to D Foley and M Dean for highlighting relevance of the maximum entropy principle in behavioural economics. DS Tourigny is a Simons Foundation Fellow of the Life Sciences Research Foundation.
      
\appendix

\section{Derivation of maximum entropy control}
\label{maxent}
To solve the optimal control problem (\ref{linprob}) one introduces the Hamiltonian
\begin{equation*}
\mathcal{H}(\Delta \mathbf{X},\mathbf{u},\boldsymbol{\lambda}) = \boldsymbol{\lambda}^T[ \mathbf{F}(\mathbf{X}(t),\mathbf{u}^0)  + \mathbf{A} \Delta \mathbf{X} + \mathbf{B} \Delta \mathbf{u} ] + \sigma H(\mathbf{u}) ,
\end{equation*}
where $\boldsymbol{\lambda}$ is a co-state vector the same dimension as $\mathbf{X}$. Applying Pontryagin's maximum principle implies maximisation of $\mathcal{H}$, or equivalently the functional
\begin{equation*}
\mathcal{F}(\mathbf{u}) =  \boldsymbol{\lambda}^T \mathbf{B} \mathbf{u} + \sigma H(\mathbf{u}) ,
\end{equation*}
with respect to $\mathbf{u}$ subject to the constraint
\begin{equation*}
\sum_{k=1}^K u_k = 1 .
\end{equation*}
This results in the $K$ first-order conditions
\begin{equation*}
\boldsymbol{\lambda}^T \mathbf{B}^k -\sigma ( 1+ \log u_k) - \alpha = 0
\end{equation*}
where $\alpha \geq 0$ is a Lagrange multiplier and $\mathbf{B}^k$ is the $k$th column of $\mathbf{B}$. The general solution of these equations takes the form
\begin{equation*}
u_k = \frac{1}{Q}\exp(\boldsymbol{\lambda}^T \mathbf{B}^k/\sigma)
\end{equation*}
and $Q$ is determined from the above constraint such that
\begin{equation*}
Q = \sum_{k=1}^K \exp(\boldsymbol{\lambda}^T \mathbf{B}^k/\sigma) .
\end{equation*}
The value of the co-state vector $\boldsymbol{\lambda}$ is obtained as in \cite{Young07} by solving the boundary value problem
\begin{equation*}
- \frac{d}{d \tau} \boldsymbol{\lambda} = \frac{\partial \mathcal{H}}{\partial \Delta \mathbf{X}} = \mathbf{A}^T \boldsymbol{\lambda} , \quad \boldsymbol{\lambda}(t + \Delta t ) = \mathbf{q}
\end{equation*} 
whose solution is 
\begin{equation*}
\boldsymbol{\lambda}(t + \tau ) = \mathbf{e}^{\mathbf{A}^T(\Delta t - \tau)} \mathbf{q} , \quad 0 \leq \tau \leq \Delta t.
\end{equation*} 
This expression for $\boldsymbol{\lambda}$ is substituted into the above expression for $u_k$ and because, ultimately, only the optimal control input at the current time $t$ is of interest, one can set $\tau = 0$ as in \cite{Young07}, which yields the maximum entropy control (\ref{control}). 

\section{First-order correction to return-on-investment}
\label{correction}
Expressed in terms of the average zeroth-order return-on-investment 
\begin{equation*}
\bar{R}_0(\mathbf{m}_s) = \sum_{k=1}^K r_k(\mathbf{m}_s)\mathbf{c}^T \mathbf{Z}^k u^0_k ,
\end{equation*}
the vector $\mathbf{q}^T\mathbf{A}$ is 
\begin{equation*}
\mathbf{q}^T\mathbf{A} = (\mathbf{0},1) \frac{\partial}{\partial \mathbf{X}} \mathbf{F}(\mathbf{X}(t),\mathbf{u}^0)= \left(x \left( \frac{\partial \bar{R}_0}{\partial \mathbf{m}_{s}} (\mathbf{m}_{s}) \right)^T,  \bar{R}_0(\mathbf{m}_s) \right) .
\end{equation*}
The first-order expansion of $\mathbf{e}^{\mathbf{A} \Delta t}$ takes the form $\mathbf{I} +  \Delta t \mathbf{A}$, where $\mathbf{I}$ is the identity matrix, and therefore the effective return-on-investment $\mathcal{R}^k_{\Delta t}(\mathbf{m}_s)$ in (\ref{eroi}) is approximated to first-order by
\begin{equation*}
x r_k(\mathbf{m}_s)\mathbf{q}^T(\mathbf{I} +  \Delta t \mathbf{A}) \begin{pmatrix}  \mathbf{S}_s  \\ \mathbf{c}^T  \end{pmatrix}  \mathbf{Z}^k .
\end{equation*}
Taking the coefficient of $x \Delta t$ and substituting for $\mathbf{q}^T\mathbf{A}$ gives the first-order correction to return-on-investment
\begin{equation*}
R^k_1(\mathbf{m}_s) = r_k(\mathbf{m}_s) \left[ \bar{R}_0(\mathbf{m}_s) \mathbf{c}^T + x \left( \frac{\partial \bar{R}_0}{\partial \mathbf{m}_{s}} (\mathbf{m}_{s}) \right)^T \mathbf{S}_s \right] \mathbf{Z}^k ,
\end{equation*} 
which can be written as 
\begin{equation*}
R^k_1(\mathbf{m}_s) = \bar{R}_0(\mathbf{m}_s) R^k_0(\mathbf{m}_s) + x Y_k(\mathbf{m}_s)
\end{equation*}
using the definition of $Y_k(\mathbf{m}_s)$ provided in (\ref{troi1}).

\section{Expressing $\mathcal{F}(\mathbf{u})$ in terms of EFM families}
\label{families}
Using (\ref{bk}) to substitute for $\mathbf{B}^k$ in the effective return-on-investment (\ref{eroi}) means that $\mathcal{F}(\mathbf{u})$ takes the form
\begin{equation*}
\mathcal{F}(\mathbf{u}) = x \mathbf{q}^T \mathbf{e}^{\mathbf{A} \Delta t} \begin{pmatrix}  \mathbf{S}_{ex}  \\ \mathbf{c}^T  \end{pmatrix} \sum_{k=1}^K r_k(\mathbf{m}_s) \mathbf{Z}^k u_k + \sigma H(\mathbf{u}) .
\end{equation*}  
The sum over $k$ in the first term can be written as
\begin{equation*}
\sum_{k=1}^K r_k(\mathbf{m}_s) \mathbf{Z}^k u_k = \sum_{J=1}^M \sum_{j \in F_J} r_j(\mathbf{m}_s) \mathbf{Z}^j u_j = \sum_{J=1}^M \left( \sum_{j \in F_J} r_j(\mathbf{m}_s) \mathbf{Z}^j \tilde{u}_j \right) U_J,
\end{equation*}
where the second equality follows from the identity $\tilde{u}_j = u_j /U_J$ for $j \in F_J$, which implies
\begin{equation*}
\mathcal{F}(\mathbf{u}) = \sum_{J=1}^M \left( \sum_{j \in F_J} \mathcal{R}^j_{\Delta t}(\mathbf{m}_s) \tilde{u}_j \right) U_J + \sigma H(\mathbf{u}) .
\end{equation*} 
Using the composition property (\ref{composition}) of entropy $H(\mathbf{u})$, one has 
\begin{equation*}
\mathcal{F}(\mathbf{u}) = \sum_{J=1}^M \left( \sum_{j \in F_J} \mathcal{R}^j_{\Delta t}(\mathbf{m}_s) \tilde{u}_j  + \sigma H(\mathbf{\tilde{u}}_J)\right) U_J + \sigma H(\mathbf{U}) 
\end{equation*}
which is (\ref{newobjf}) with 
\begin{equation*}
\mathcal{F}_J(\mathbf{\tilde{u}}_J) = \sum_{j \in F_J} \mathcal{R}^j_{\Delta t}(\mathbf{m}_s) \tilde{u}_j  + \sigma H(\mathbf{\tilde{u}}_J)
\end{equation*}
the objective functional $\mathcal{F}$ restricted to the $J$th family. Notice that the weighting (\ref{weighting}) defines the effective return-on-investment for the $J$th family derived directly from system (\ref{reduced2}) to be 
\begin{equation*}
\mathcal{\tilde{R}}^J_{\Delta t} (\mathbf{m}_s) = \sum_{j \in F_J}  \mathcal{R}^j_{\Delta t}(\mathbf{m}_s) \tilde{u}_j 
\end{equation*}
and therefore $\mathcal{F}_J(\mathbf{\tilde{u}}_J)$ includes an entropic correction that is only zero when $\tilde{u}_J$ describes an FBA/Bang-Bang policy.

\end{document}